\newif{\ifarxiv}
\newif{\ifdraft}
\newif{\ifremarks}

\arxivtrue  
\remarkstrue  

\documentclass[11pt,a4paper]{article}
\pdfoutput=1
\usepackage[utf8]{inputenc}
\usepackage[left=2.8cm,right=2.8cm,top=2.8cm,bottom=2.8cm]{geometry}
\usepackage[nottoc]{tocbibind} 
\usepackage{graphicx}
\usepackage{amsmath,amssymb,mathtools,shuffle}
\ifdraft\usepackage{showkeys}\fi 
\usepackage[bookmarks=true,ocgcolorlinks=true]{hyperref} 
\usepackage{cite}
\usepackage[bulletsep]{collref} 
\usepackage{xcolor}
\usepackage{booktabs}

\usepackage{lmodern}
\usepackage[T1]{fontenc}
\usepackage{tgpagella}
\usepackage{mathpazo}
\usepackage{microtype}

\usepackage{pifont}\collectsep{\ding{72}~}   

\ifremarks
\newcommand{\remark}[1]{{\renewcommand{\bfdefault}{b}{\color[RGB]{0,150,0}{\textbf{#1}}}}}
\fi
\providecommand{\remark}[1]{\ignorespaces}

\usepackage[font=small,labelfont=bf,width=0.85\textwidth]{caption}

\providecommand{\hypersetup}[1]{}
\providecommand{\hyperref}[2][]{#2}
\providecommand{\texorpdfstring}[2]{#1}
\providecommand{\pdfbookmark}[3][]{}

\hypersetup{plainpages=false}
\hypersetup{pdfpagemode=UseOutlines}
\hypersetup{bookmarksnumbered=true}
\hypersetup{bookmarksopen=true}
\hypersetup{pdfstartview=FitH}
\hypersetup{citecolor=[rgb]{0 .4 0}}
\hypersetup{urlcolor=[rgb]{.4 0 0}}
\hypersetup{linkcolor=[rgb]{0 0 .4}}
\hypersetup{citebordercolor={.6 .9 .6}}
\hypersetup{urlbordercolor={.7 .8 1}}
\hypersetup{linkbordercolor={1 .7 .7}}
\hypersetup{pdfborder={0 0 .5}}

\makeatletter
\renewcommand*\l@section[2]{%
  \ifnum \c@tocdepth >\z@
    \addpenalty\@secpenalty
    \addvspace{0.2em \@plus\p@}%
    \setlength\@tempdima{1.5em}%
    \begingroup
      \parindent \z@ \rightskip \@pnumwidth
      \parfillskip -\@pnumwidth
      \leavevmode \bfseries
      \advance\leftskip\@tempdima
      \hskip -\leftskip
      #1\nobreak\hfil \nobreak\hb@xt@\@pnumwidth{\hss #2}\par
    \endgroup
  \fi}
\renewcommand\tableofcontents{%
    \@starttoc{toc}%
    }
\makeatother

\newcommand{\namedref}[2]{\hyperref[#2]{#1~\ref*{#2}}}
\newcommand{\secref}[1]{\namedref{Section}{#1}}
\newcommand{\appref}[1]{\namedref{Appendix}{#1}}
\newcommand{\tabref}[1]{\namedref{Table}{#1}}
\newcommand{\figref}[1]{\namedref{Figure}{#1}}

\makeatletter
\def\mr@ignsp#1 {\ifx\:#1\@empty\else #1\expandafter\mr@ignsp\fi}%
\newcommand{\multiref}[1]{\begingroup
\xdef\mr@no@sparg{\expandafter\mr@ignsp#1 \: }%
\def\mr@comma{}%
\@for\mr@refs:=\mr@no@sparg\do{\mr@comma\def\mr@comma{,}\ref{\mr@refs}}%
\endgroup}
\makeatother
\renewcommand{\eqref}[1]{(\multiref{#1})}

\makeatletter
\let\@myabstract\@empty
\let\@keywords\@empty
\let\@subject\@empty
\providecommand{\affiliation}[1]{\gdef\@affiliation{#1}}
\providecommand{\myabstract}[1]{\gdef\@myabstract{#1}}
\providecommand{\keywords}[1]{\gdef\@keywords{#1}}
\providecommand{\subject}[1]{\gdef\@subject{#1}}
\def\thetitle{\@title}
\def\theauthor{\@author}
\def\theaffiliation{\@affiliation}
\def\theabstract{\@myabstract}
\def\thesubject{\@subject}
\def\thedate{\@date}
\def\thekeywords{\@keywords}
\makeatother
\def\fillpdfdata{
\hypersetup{pdftitle={\thetitle}}%
\hypersetup{pdfsubject={\thesubject}}%
\hypersetup{pdfkeywords={\thekeywords}}%
}

\let\oldbfseries=\bfseries
\let\oldmdseries=\mdseries
\let\oldnormalfont=\normalfont
\renewcommand{\bfseries}{\oldbfseries\boldmath}
\renewcommand{\mdseries}{\oldmdseries\unboldmath}
\renewcommand{\normalfont}{\oldnormalfont\unboldmath}

\numberwithin{equation}{section}

\makeatletter
\newlength{\apb@width}
\newcommand{\autoparbox}[2][c]{\settowidth{\apb@width}{#2}\parbox[#1]{\apb@width}{#2}}
\newcommand{\includegraphicsbox}[2][]{\autoparbox{\includegraphics[#1]{#2}}}
\makeatother

\newcommand{\nn}{\nonumber}

\newcommand{\brk}[1]{(#1)}
\newcommand{\lrbrk}[1]{\left(#1\right)}
\newcommand{\bigbrk}[1]{\bigl(#1\bigr)}

\newcommand{\Bigbrk}[1]{\Bigl(#1\Bigr)}
\newcommand{\sbrk}[1]{[#1]}

\newcommand{\biggsbrk}[1]{\biggl[#1\biggr]}
\newcommand{\Bigsbrk}[1]{\Bigl[#1\Bigr]}
\newcommand{\brc}[1]{\{#1\}}

\newcommand{\bigbrc}[1]{\bigl\{#1\bigr\}}

\newcommand{\sfrac}[2]{{\textstyle\frac{#1}{#2}}}
\newcommand{\half}{\sfrac{1}{2}}

\newcommand{\p}{{+}}
\newcommand{\m}{{-}}
\newcommand{\superN}{\mathcal{N}}
\newcommand{\grp}[1]{\mathrm{#1}}

\newcommand{\dlog}{\operatorname{dlog}}
\newcommand{\Li}{\operatorname{Li}}
\newcommand{\abs}[1]{|#1|}
\newcommand{\smbop}[1]{\operatorname{S}\sbrk{#1}}

\newcommand{\Z}{\mathbb{Z}}
\newcommand{\C}{\mathbb{C}}
\newcommand{\order}[1]{\mathcal{O}(#1)}
\newcommand{\amp}{\mathcal{A}}

\newcommand{\supMRL}{^{\mathrm{MRL}}}
\newcommand{\supMHV}{^{\mathrm{MHV}}}
\newcommand{\suprm}[1]{^{\text{#1}}}
\newcommand{\subrm}[1]{_{\text{#1}}}
\newcommand{\dd}{\mathrm{d}}
\newcommand{\Nc}{N\subrm{c}}
\newcommand{\Gs}{G^{\mathbf{s}}}
\newcommand{\Gsshort}[1]{G^{\mathbf{s},#1}}
\newcommand{\lyndon}{\operatorname{Lyn}}
\newcommand{\id}{\operatorname{id}}
\newcommand{\cx}{\check x}
\newcommand{\cy}{\check y}

\newcommand{\bfm}[1]{\boldsymbol{#1}}

\newcommand{\mathematica}{\textsc{Mathematica}}
\newcommand{\filename}[1]{\texttt{#1}}

\begin{document}

\ifdraft
\overfullrule=5pt
\fi


\title{Systematics of the Multi-Regge Three-Loop Symbol}

\myabstract{We review the systematics of Mandelstam cut contributions
to planar scattering amplitudes
in the multi-Regge limit. Isolating the relevant cut terms, we explain
how the BFKL expansion can be
used to construct the perturbative $n$-point
multi-Regge limit amplitude in certain kinematic regions from a finite number
of basic building blocks. At three loops and at leading
logarithmic order, two building
blocks are required. Their symbols are extracted from the known three-loop
six-point and seven-point symbols for general
kinematics. The new seven-point building block is constructed in terms
of single-valued multiple polylogarithms to the extent it can be
determined using the symbol as well as further symmetry and consistency constraints.
Beyond the leading logarithmic order, the subleading and sub-subleading terms
require two and one further building block, respectively. The latter
could either be reconstructed from further perturbative
data, or from BFKL integrals involving yet-unknown
corrections to the central emission block.}

\subject{Theoretical Physics}



\ifarxiv
\noindent
\mbox{}\hfill DESY 16-115
\fi

\author{%
Till Bargheer
}

\hypersetup{pdfauthor={\theauthor}}

\vfill

\begin{center}
{\Large\textbf{\mathversion{bold}\thetitle}\par}
\vspace{1cm}

\textsc{\theauthor}

\medskip

\textit{%
DESY Theory Group, DESY Hamburg
\\
Notkestra{\ss}e 85, D-22607 Hamburg, Germany
}

\medskip

{\ttfamily
\href{mailto:till.bargheer@desy.de}{till.bargheer@desy.de}
}
\par\vspace{1cm}

\textbf{Abstract}\vspace{5mm}

\begin{minipage}{12.4cm}
\theabstract
\end{minipage}

\vfill

\end{center}

\fillpdfdata

\hrule
\vspace{2ex}
\providecommand{\microtypesetup}[1]{}
\microtypesetup{protrusion=false}
\setcounter{tocdepth}{1}
\tableofcontents
\microtypesetup{protrusion=true}
\vspace{3ex}\hrule

\vfill

\newpage

\section{Introduction}

Recent advances in the study of scattering amplitudes have sparked
renewed interest in the multi-Regge limit of high-energy scattering.
Besides its phenomenological significance, it has long been noted that
the perturbative expansion simplifies considerably in this limit:
Typically, the perturbative series has to be (and in fact can be!)
resummed due to the appearance of large logarithms, leading to
factorized all-order expressions for scattering processes. A further
enhancement comes about in the case of planar $\superN=4$ super
Yang--Mills theory: Here, the multi-Reggeon states that resum
all-order gluon exchanges are governed by the integrable
Balitsky--Fadin--Kuraev--Lipatov
(BFKL)~\cite{Fadin:1975cb,Kuraev:1976ge,Balitsky:1978ic} and
Bartels--Kwieci\`nski--Prasza{\l}owicz
(BKP)~\cite{Bartels:1980pe,Kwiecinski:1980wb} Hamiltonians.
This first appearance of integrability in the planar theory was
observed long before the extensive discoveries and applications of
integrable structures that took place during the past fifteen
years~\cite{Beisert:2010jr}. Since the proposal of the exponentiated
Bern--Dixon--Smirnov (BDS) amplitude~\cite{Bern:2005iz}, the
systematics of multi-Regge limit
amplitudes in planar $\superN=4$ super Yang--Mills theory have
been understood to a remarkable extent. In fact, after a disagreement
at strong coupling had casted doubt on the correctness of the BDS
amplitude~\cite{Alday:2007he}, it was the absence of
the expected Regge pole and cut terms that invalidated the proposal at
weak coupling~\cite{Bartels:2008ce}, and that prompted the correction
of the
BDS amplitude by the dual conformally invariant remainder function
beyond five points~\cite{Bartels:2008sc,Lipatov:2010qf}.

By now, the remainder function has been constructed to high loop
orders by constraining the possible function space through physical
symmetry and analyticity
requirements~\cite{Dixon:2014voa,Dixon:2015iva}. This \emph{bootstrap program}
relies on various input, ranging from the mathematical theory of the
relevant functions~\cite{Goncharov:2001iea,BROWN2004527}
to recursion
relations~\cite{CaronHuot:2011kk} and
the expansion around collinear limits as dictated by
integrability~\cite{Basso:2013vsa,Basso:2013aha}.
In all cases, knowledge about the multi-Regge limit has provided
important boundary data to the bootstrap enterprise. Conversely, these
recent methods admit to compute the BFKL data, and hence the
multi-Regge-limit remainder function, to unprecedented
orders~\cite{Dixon:2012yy,Dixon:2013eka,Dixon:2014voa,Drummond:2015jea,Broedel:2015nfp,Broedel:2016kls}.
To date, this fruitful interplay has mostly been restricted to the
six-point case. An extension to seven-point functions has been
initiated recently~\cite{Drummond:2014ffa}. Going to even higher points will require a better
understanding of the relevant function space. It is conceivable that
the Regge limit will again provide valuable boundary data in this regard.

It has been understood that obtaining the full analytic structure of multi-Regge
limit amplitudes requires to analyze the amplitudes in all possible
kinematic
regions~\cite{Bartels:2011xy,Bartels:2011ge,Bartels:2013jna,Bartels:2014jya}.
In fact, while the integrable structure at strong coupling becomes
particularly amenable in the multi-Regge limit~\cite{Bartels:2012gq}, a discrepancy with
the expectation from weak coupling has been
observed in one of the kinematic regions at seven points~\cite{Bartels:2014mka}.
Recently, a systematic study of the $n$-point two-loop
remainder function in all kinematic regions at weak coupling has been put
forward~\cite{Bargheer:2015djt}. The ability to study any number of
points relied on the known
two-loop \emph{symbol} of the remainder
function for all multiplicities~\cite{CaronHuot:2011ky}. Passing from
polylogarithmic functions to their symbols constitutes a major
simplification, both for the analysis of the relevant expressions and
for the systematics of the multi-Regge limit.

The goal of the present work is two-fold: One aim is to understand the
results of the previous study~\cite{Bargheer:2015djt} from the
perspective of Regge cut contributions. Secondly, we want to lift the
analysis to the three-loop level. To this end, we first isolate the Regge
cut contributions that contribute to a given region, and then expand
the relevant contributions to the three-loop order. Judiciously
grouping the resulting terms, we find that the $n$-point three-loop
remainder function, in the simplest class of kinematic regions, reduces to a linear
combination of five building block functions. At the symbol
level, the reduction extends to all kinematic regions. The
symbols of the two building blocks required to reconstruct the $n$-point
remainder function at leading logarithmic order are extracted from
the known perturbative data. To the extent that it is fixed by the
symbol as well as symmetries, the new seven-point building block is
constructed in terms of multiple polylogarithms.
The results of this work are assembled in a
computer-readable file attached to this submission.

\paragraph{Overview.}

\secref{sec:background} briefly summarizes the systematics of
planar scattering amplitudes in the multi-Regge limit in a
self-contained way. \secref{sec:symbols} highlights the
simplifications and restrictions implied by specializing to certain
kinematic regions, or by passing from functions to
symbols. In \secref{sec:2loop}, the two-loop analysis
of the multi-Regge limit remainder function is revisited from the Regge cut point
of view. \secref{sec:3loop} extends the analysis to three loops,
where the remainder function can be decomposed into a few
basic building blocks. The latter are discussed in
\secref{sec:blocks}. We construct a function for the seven-point
building block in \secref{sec:g3function}, and \secref{sec:concl} presents the conclusion.

\paragraph{Note added:}

The simultaneous paper~\cite{DelDuca:2016lad} has some overlap with
the present work. In particular, there is a connection between the
``factorization theorem'' of~\cite{DelDuca:2016lad} and the
application of the reduction identities~\eqref{eq:reduction} to the
expansion of the $n$-point cut contribution carried out in this work.

\section{Background}
\label{sec:background}

\paragraph{Multi-Regge Kinematics.}

\begin{figure}\centering
\includegraphics{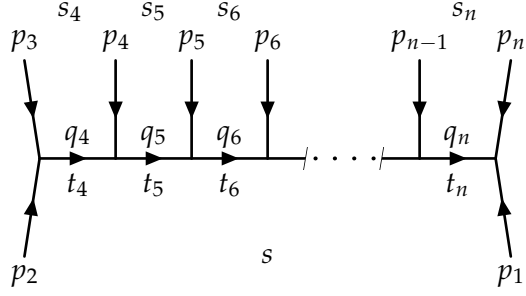}
\caption{Kinematic variables.}
\label{fig:kinematics}
\end{figure}

The $2\to(n-2)$ multi-Regge limit is the $n$-particle generalization
of the simple $s\gg t$ Regge limit for $2\to2$ scattering. To describe a general
amplitude,  we will use the $(3n-10)$ independent Lorentz invariants
\begin{gather}
t_j\equiv \bfm{q}_j^2
\,,\qquad
\bfm{q}_j\equiv \bfm{p}_2+\bfm{p}_3+\dots+\bfm{p}_{j-1}
\,,\qquad
j=4,\dots,n
\,,\\
s_j\equiv s_{j-1,j}\equiv(\bfm{p}_{j-1}+\bfm{p}_j)^2
\,,\qquad
j=4,\dots,n
\,,\\
\eta_j\equiv\frac{s_js_{j+1}}{(\bfm{p}_{j-1}+\bfm{p}_j+\bfm{p}_{j+1})^2}
\,,\qquad j=4,\dots,n-1
\,.
\end{gather}
Here, $\bfm{p}_1,\dots,\bfm{p}_n$ are the $n$ external momenta. By convention,
they are all incoming, but may have either energy sign.
The $2\to(n-2)$ multi-Regge limit is characterized by a large
separation of rapidities among the produced particles. In terms of the
above kinematic variables, the limit is attained for
\begin{equation}
\abs{s}\gg\abs{s_4},\dots,\abs{s_n}\gg t_4,\dots,t_n
\,,
\end{equation}
where $s=(\bfm{p}_1+\bfm{p}_2)^2$ is the total energy. See \figref{fig:kinematics}
for an illustration of the kinematics.
Many quantities in the multi-Regge limit only depend on the kinematics
in the transverse space to the $(\bfm{p}_1,\bfm{p}_2)$ plane. We hence define
\begin{equation}
\bfm{p}_j=\alpha_j \bfm{p}_1+\beta_j \bfm{p}_2+\bfm{p}_{j}^{\perp}
\,,\qquad
\bfm{p}_1\cdot\bfm{p}_{j}^{\perp}=\bfm{p}_2\cdot\bfm{p}_{j}^{\perp}=0
\,,\qquad
j=4,\dots,n-1
\,,
\end{equation}
and similarly for $\bfm{q}_4,\dots,\bfm{q}_n$. It is often convenient
to switch to complex variables $p_j$, $q_j$ whose real and imaginary
parts equal the two components of the transverse momenta
$\bfm{p}_{j}^{\perp}$ and $\bfm{q}_{j}^{\perp}$, respectively:
\begin{equation}
\bfm{p}_{j}^{\perp}=\bigbrk{\Re(p_j),\Im(p_j)}
\,,\qquad
\bfm{q}_{j}^{\perp}=\bigbrk{\Re(q_j),\Im(q_j)}
\,.
\end{equation}
Frequently used combinations of the transverse momenta are the complex anharmonic
ratios
\begin{equation}
w_j=\frac{p_{j-1}q_{j+1}}{q_{j-1}p_j}
\,,\qquad
j=5,\dots,n-1
\,.
\label{eq:wj}
\end{equation}
Planar $\superN=4$ super Yang--Mills
theory enjoys dual conformal invariance. Invariant quantities in
this theory can thus only depend on conformally invariant cross ratios
\begin{equation}
U_{ij}\equiv\frac{x_{i+1,j}^2x_{i,j+1}^2}{x_{ij}^2x_{i+1,j+1}^2}
\,,\qquad
3\leq\abs{i-j}\leq n-2
\label{eq:Uij}
\end{equation}
of the dual coordinates
\begin{equation}
\bfm{p}_j\equiv x_j-x_{j-1}
\,,\qquad
x_{ij}=x_i-x_j
\,.
\end{equation}
A basis of kinematically independent invariant cross ratios is
provided by
\begin{equation}
u_{j,1}=U_{j-3,j}
\,,\qquad
u_{j,2}=U_{j-2,n}
\,,\qquad
u_{j,3}=U_{1,j-1}
\,,\qquad
j=5,\dots,n-1
\,.
\label{eq:uj}
\end{equation}
In the multi-Regge limit, these cross ratios converge to $1$ or $0$:
\begin{equation}
u_{j,1}\to1
\,,\qquad
u_{j,2}\to0
\,,\qquad
u_{j,3}\to0
\,.
\end{equation}
The ratios of subleading terms remain finite, and are related to the
anharmonic ratios~\eqref{eq:wj} via
\begin{equation}
\frac{u_{j,2}}{1-u_{j,1}}\to\frac{1}{\abs{1+w_{j}}^2}
\,,\qquad
\frac{u_{j,3}}{1-u_{j,1}}\to\frac{\abs{w_{j}}^2}{\abs{1+w_{j}}^2}
\,.
\label{eq:uofw}
\end{equation}
%

\paragraph{Kinematic Regions.}

In order to understand the full analytic structure of the multi-Regge
limit amplitude, it is important to analyze it in all
physical kinematic regions. Our starting point will be the physical region in
which the energies of all particles $3,\dots,n$ are negative (which
means that those particles are effectively outgoing, instead of
incoming).
In this region, all subenergies $s_j$, $j=4,\dots,n$, are
negative.%
\footnote{The Minkowski metric is assumed to have signature $\brk{\m\p\p\p}$.}
In all other physical regions that we will consider, some of the
particles $4,\dots,n-1$ have positive energies (those particles become
incoming), and hence some of the
invariants $s_j$ become positive. These other regions are sometimes called
``Mandelstam regions''.%
\footnote{One could also consider regions in which the energies of
particles $3$ and/or $n$ are positive, but those regions do not add
further analytic structure to the amplitude, and will thus not be
considered in the following.}
They can be reached from the
all-outgoing region by analytic continuation of the kinematics.
The various regions will be labeled by the subsets
$I\subset\brc{4,\dots,n-1}$ of particles whose energies have been
continued to positive values.
Alternatively, we will often label
regions by $\rho=\brk{\rho_4,\dots,\rho_{n-1}}\in\Z_2^{n-4}$, with
$\rho_j=\pm1$ (or just $\rho_j=\pm$) indicating whether the
respective particle has been flipped (its energy has been continued) $(\m)$ or not
$(\p)$.

Importantly, the various regions become disconnected in the strict
multi-Regge limit. That is to say, in order to continue the kinematics
from one region to the other, one has to complexify the subenergies
$s_k$ (e.g.\ by continuing them along big circles).

\paragraph{Multi-Regge Limit Amplitudes.}

Scattering in the multi-Regge limit is dominated by the exchange of
``Reggeized gluons'' (or ``Reggeons''), which are effective particles
that resum the contributions of entire classes of gluonic Feynman
diagrams of all loop orders. The simplest example is the four-point
amplitude in \emph{planar} $\superN=4$ super Yang--Mills theory,
for which
\emph{all} perturbative contributions can be resummed and factorized into a single
diagram:
\begin{equation}
\amp_4\supMRL
=\includegraphicsbox{Fig4pt}
=\Gamma(t)s^{\omega(t)}\Gamma(t)
\,.
\end{equation}
Here, $\includegraphics{FigGGRvertex}=\Gamma(t)$ is the gluon-gluon-Reggeon
vertex (see e.g.~\cite{Bartels:2008ce}), and
\includegraphicsbox{FigReggeon} stands for the exchange of a single
Reggeon with propagator $s^{\omega(t)}$, where
$\omega(t)$ is the (real-valued) Regge trajectory.
At five points, two different kinematic regions can be considered:
The produced particle $4$ can either be flipped $(\m)$ or
not $(\p)$.
Strikingly, the factorization property of the
four-point amplitude extends to this case: In both regions, the
planar five-point amplitude again factorizes into a single diagram,
\begin{equation}
\amp_5^{\mathrm{MRL}(\pm)}
=\includegraphicsbox{Fig5pt}
=
\Gamma(t_4)
\,s_4^{\omega(t_4)}
\,\Gamma_{45}
\,s_5^{\omega(t_5)}
\,\Gamma(t_5)
\,,
\label{eq:A5}
\end{equation}
where the
(complex) gluon production vertex~\cite{Lipatov:2010qf}
\begin{equation}
\includegraphicsbox{FigRRGvertex}
=\Gamma_{j,j+1}
=\abs{\Gamma_{j,j+1}}e^{\pm i\pi\,\hat\omega_{j,j+1}}
\,,\qquad
\hat\omega_{j,j+1}=\hat\omega(t_j,t_{j+1},\eta_j)
\,,
\label{eq:RRGvertex}
\end{equation}
only depends on the kinematic region through the sign of its phase.

A general $n$-point multi-Regge-limit amplitude in any given kinematic region
$\rho$ receives contributions from
Regge pole~\cite{Regge:1960zc} as well as Mandelstam cut terms~\cite{Amati:1962nv,Mandelstam:1963cw},
\begin{equation}
\amp_{n,\rho}\supMRL
=\amp_{n,\rho}\suprm{Regge pole}
+\amp_{n,\rho}\suprm{Mandelstam cut}
\,.
\end{equation}
Both the pole terms and the cut terms depend on the kinematic region
$\rho$.
The origin of the Mandelstam cut terms are non-trivial
contributions from multi-Reggeon bound state exchange in intermediate
$t$-channels. For planar amplitudes of up to five points, such
contributions are suppressed by powers of $1/\Nc$, and the
amplitudes factorize as indicated above. At six points, the
first cut term appears, in the region $(\m\m)$ where both intermediate
momenta have been flipped~\cite{Bartels:2008sc}.
In a generic region $\rho$, the six-point amplitude therefore
reads~\cite{Lipatov:2010qf,Bartels:2013jna}
\begin{equation}
\amp_{6,\rho}\supMRL
=\text{(pole terms)}^\rho
+c_{6,1,4}^\rho\,\includegraphicsbox{Fig6pt2}
\,,
\label{eq:A6full}
\end{equation}
where the region-dependent coefficient $c_{6,1,4}^\rho$ is
non-vanishing for $\rho=(\m\m)$.
Here, the cut diagram stands for all contributions from two-Reggeon
bound state exchange in the $t_5$ channel.
This picture generalizes to higher multiplicities: The planar $n$-point
multi-Regge limit amplitude is a sum of region-dependent Regge pole
terms as well as Mandelstam cut
contributions
with region-dependent
coefficients~\cite{Bartels:2008ce,Bartels:2008sc,Lipatov:2010qf,Lipatov:2010ad,Bartels:2011ge,Bartels:2013jna,Bartels:2014jya,Bartels:8points}:
\begin{multline}
\amp_{n,\rho}\supMRL
=\text{(pole terms)}^\rho
+\sum_jc_{n,1,j}^\rho\includegraphicsbox{Fignpt2}
+\sum_jc_{n,2,j}^\rho\includegraphicsbox{Fignpt22}
+\dots
\\
+\sum_jd_{n,1,j}^\rho\includegraphicsbox{Fignpt232}
+\sum_{j<k}e_{n,j,k}^\rho\includegraphicsbox{Fignpt212}
+\dots
\label{eq:Anfull}
\end{multline}
Here, the symbol \includegraphicsbox{FigBreak} stands for the insertion of zero or more
complex gluon production vertices~\eqref{eq:RRGvertex}.
For planar amplitudes, the number of exchanged Reggeons can at most
increase or decrease by one when passing from one $t$-channel to the
next.%
\footnote{The number $M_n$ of admissible diagrams that can contribute
to the $n$-point amplitude, as a sequence in $n$, equals the Motzkin
sequence, \href{http://oeis.org/A001006}{OEIS A001006}~\cite{OEIS},
with $M_n/M_{n-1}\to3$ for $n\to\infty$.}
All other contributions are suppressed by powers of $1/\Nc$.
The pole terms as well as the cut-term
prefactors can in principle be obtained from the general quantum field
theory principles of locality \& unitarity. The procedure particularly
relies on
an expansion of the amplitude into
a sum of terms that each
have no overlapping energy discontinuities, following the Steinmann
relations~\cite{Steinmann:1960a}. Determining the cut
contributions in this way is a very
intricate and tedious procedure that has to be carried out region by
region. This formidable task has been completed for the seven-point
amplitude~\cite{Bartels:2014jya}, and a study of the eight-point case
is underway~\cite{Bartels:8points}, but a generalization to higher
multiplicities appears difficult.
Below, we will see that the coefficients $c_{n,b,j}$ are actually
fixed by the two-loop analysis~\cite{Bargheer:2015djt}.

In fact, the \emph{Mandelstam criterion}~\cite{Mandelstam:1963cw}
significantly constrains the set of
cut terms that can contribute to any given kinematic region: It
asserts that any cut contribution in which the multi-Reggeon states
span the adjacent $t$-channels $t_j,\dots,t_k$
cannot contribute to regions in which $s_{j-1}>0$ or $s_{k+1}>0$, that is%
\footnote{The reason is that in such cases, one of the Feynman loop
integrals can be closed trivially, since all
singularities lie on the same side of the integration
contour~\cite{Lipatov:2009nt}.}
\begin{equation}
c_{n,k-j,j}^\rho=0
\qquad
\text{if}
\qquad
\rho_{j-1}=\rho_j
\quad
\text{or}
\quad
\rho_k=\rho_{k+1}
\,,
\label{eq:Mandelstamcrit}
\end{equation}
and similarly for the further coefficients in~\eqref{eq:Anfull}.
Here, the subscripts $n$, $b$, and $j$ in $c_{n,b,j}^{\rho}$
label the total number of particles, the number of $t$-channels taking
part in the multi-Reggeon state, and the produced gluon that bounds
the multi-Reggeon state on the left. For example,
as indicated above, the six-particle cut term~\eqref{eq:A6full} is
only present in the $(\m\m)$ region:
\begin{equation}
c_{6,1,4}^{(\p\p)}=c_{6,1,4}^{(\p\m)}=c_{6,1,4}^{(\m\p)}=0\,.
\end{equation}
%

\paragraph{BDS and Remainder Function.}

The MHV amplitudes of planar $\superN=4$ super Yang--Mills theory
can be decomposed into
two factors:
\begin{equation}
\amp_n\supMHV
=
\amp_n\suprm{BDS}
R_n
\end{equation}
Here,
$\amp_n\suprm{BDS}$ is the
Bern--Dixon--Smirnov amplitude~\cite{Bern:2005iz}, wich equals
the tree-level amplitude times
the exponentiated one-loop
amplitude, and which in fact produces the correct all-loop four-point and
five-point amplitudes. Starting at six points, it however fails to
reproduce the correct Regge pole contributions, and it misses all
Regge cut terms (beyond one
loop)~\cite{Bartels:2008sc,Bartels:2008ce,Lipatov:2010qf}.
Hence it cannot be the full
amplitude, but has to be corrected by a non-trivial \emph{remainder function}
$R_n$. Since the BDS amplitude correctly captures all
infrared singularities and dual conformal
weights, the remainder function is infrared finite and dual
conformally invariant,
and thus can only depend on dual conformally
invariant cross ratios~\eqref{eq:Uij}. By definition, it is only non-trivial starting from six points and
two loops.

Passing to the multi-Regge limit, and stripping off the universal
absolute value, the BDS amplitude reduces to a region-dependent phase factor. From the
latter, one can separate off a conformally invariant, infrared finite
part $\exp(i\delta_n^\rho)$, which again is region-dependent, and
contains the finite part of the one-loop Regge cut
terms~\cite{Lipatov:2010qf,Bartels:2013jna}
\begin{equation}
\frac{\amp_n^{\mathrm{BDS,MRL},\rho}}{\Gamma(t_4)\,\abs{s_4^{\omega_4}}\,\abs{\Gamma_{45}}\,\abs{s_5^{\omega_5}}\,\abs{\Gamma_{56}}\,\abs{s_6^{\omega_6}}\dots
\abs{s_{n-1}^{\omega_{n-1}}}\,\abs{\Gamma_{n-1,n}}\,\abs{s_n^{\omega_n}}\Gamma(t_{n})}
=\exp\brk{i\phi_n^\rho}\exp\brk{i\delta_n^\rho}
\,.
\end{equation}
The universal denominator is a generalization of the five-point
amplitude~\eqref{eq:A5}, and it subsumes all dependence on the
absolute values of the gluon production
vertices $\Gamma_{k,k+1}$ and Reggeon propagators $s_k^{\omega_k}$.
The region-dependent phase $\exp\brk{i\phi_n^\rho}$ absorbs the remaining infrared
divergences.
The finite, conformally invariant piece $\exp\brk{i\delta_n^\rho}$ combines in a non-trivial
way with the
remainder function
to a region-dependent
linear combination of reduced pole and cut
terms~\cite{Bartels:2013jna,Bartels:2014jya}:
\begin{multline}
\exp\brk{i\delta_n^\rho}R_n^\rho
=\text{(reduced pole terms)}^\rho
\\
+\sum_jc_{n,1,j}^\rho\includegraphicsbox{Fignpt2r}
+\sum_jc_{n,2,j}^\rho\includegraphicsbox{Fignpt22r}
+\dots
\\
+\sum_jd_{n,1,j}^\rho\includegraphicsbox{Fignpt232r}
+\sum_{j<k}e_{n,j,k}^\rho\includegraphicsbox{Fignpt212r}
+\dots
\label{eq:allcuts}
\end{multline}
Here, the grayed-out parts of the cut diagrams have been divided out,
and the (black) cut pieces stand for the remainder after the division.

\paragraph{Factorized Cut Integrals.}

All reduced cut terms in~\eqref{eq:allcuts} are infrared-finite, conformally
invariant functions of the complex anharmonic ratios
$w_k$~\eqref{eq:wj}. Just like the pole terms of the four-point and
five-point amplitudes, they enjoy the virtue of Regge factorization,
in the following sense: The multi-Reggeon bound states that propagate
in the intermediate $t$-channels are governed by the
BFKL~\cite{Fadin:1975cb,Kuraev:1976ge,Balitsky:1978ic} and
BKP~\cite{Bartels:1980pe,Kwiecinski:1980wb} equations. The solutions
to these equations are most naturally expressed in terms of their
$\grp{SL}(2,\C)$ representation labels $(n,\nu)$. Expressing all quantities in terms
of these variables, the cut contribution factorizes into a simple
product: Reading a cut diagram from left to right, each $t$-channel
$m$-Reggeon state contributes one BFKL (or BKP) \emph{Green's
function} $G_m(n_k,\nu_k)$, each gluon emission that increments or
decrements the number of exchanged reggeons from $m$ to $m\pm1$
contributes an \emph{impact factor}
$\Phi_{m,m\pm1}(n_{k-1},\nu_{k-1},n_k,\nu_k)$, and each intermediate
gluon $k$ that gets emitted from an $m$-Reggeon bound state
contributes a \emph{central emission block}
$C_{m}(n_{k-1},\nu_{k-1},n_k,\nu_k)$. Obtaining the full cut
contribution requires completing the state sums in all
$t$-channels by summing and integrating over all
$n_k$ and $\nu_k$. The summation and integration amounts to a Fourier--Mellin
transform from the $(n_k,\nu_k)$ variables to
the complex anharmonic ratios $w_k$
that provide the kinematic dependence.

The subsequent analysis will focus on the cuts of the type shown in the
middle line of~\eqref{eq:allcuts}. For those terms, only the simplest
impact factors~\cite{Bartels:2008sc}
\begin{align}
\Phi_{\mathrm{L},k} &\equiv \Phi_{0,1}(n_k,\nu_k)
=
\frac{1}{2}\frac{(-1)^n}{i\nu_k+n_k/2}
\lrbrk{\frac{q_{k-1}}{p_{k-1}}}^{-i\nu_k-n_k/2}
\lrbrk{\frac{\bar q_{k-1}}{\bar p_{k-1}}}^{-i\nu_k+n_k/2}
+\order{g}
\,,
\nn\\
\Phi_{\mathrm{R},k} &\equiv \Phi_{1,0}(n_k,\nu_k)
=
-\frac{1}{2}\frac{1}{i\nu_k-n_k/2}
\lrbrk{\frac{q_{k+1}}{p_{k}}}^{i\nu_k+n_k/2}
\lrbrk{\frac{\bar q_{k+1}}{\bar p_{k}}}^{i\nu_k-n_k/2}
+\order{g}
\label{eq:iflead}
\end{align}
and emission blocks~\cite{Bartels:2011ge}
\begin{multline}
C_k \equiv C_{1}(n_{k},\nu_{k},n_{k+1},\nu_{k+1})
=-\frac{1}{2}
\lrbrk{\frac{q_{k+1}}{p_k}}^{i\nu_{k}+n_{k}/2}
\lrbrk{\frac{\bar q_{k+1}}{\bar p_k}}^{i\nu_{k}-n_{k}/2}
\cdot\\\cdot
\lrbrk{\frac{q_{k}}{p_k}}^{-i\nu_{k+1}-n_{k+1}/2}
\lrbrk{\frac{\bar q_{k}}{\bar p_k}}^{-i\nu_{k+1}+n_{k+1}/2}
\tilde C(n_{k},\nu_{k},n_{k+1},\nu_{k+1})
+\order{g}
\label{eq:evlead}
\end{multline}
are needed. The required Green's function stems from the BFKL
color-octet channel and takes the form~\cite{Bartels:2008sc}
\begin{equation}
G_k \equiv G_2(n_k,\nu_k)
=
\varepsilon_k^{gE_{n_k,\nu_k}}
\,.
\label{eq:bfklgreen}
\end{equation}
Here,
$\varepsilon_k\equiv -\sqrt{u_{k,2}u_{k,3}}$ are combinations of
``small'' cross ratios~\eqref{eq:uj} that approach zero
in the multi-Regge limit, $E_{n,\nu}$ is the BFKL
color-octet eigenvalue, and
\begin{equation}
g\equiv\frac{g\subrm{YM}^2\Nc}{8\pi^2}
\end{equation}
is the planar coupling constant.
The general two-Reggeon cut term $f_k$ spanning $k$ $t$-channels therefore takes the
form~\cite{Bartels:2011ge,Bartels:2014mka}%
\footnote{The cut contribution is normalized such that the cut
coefficient $c_{6,1,4}^{\rho=(\m\m)}$ of the six-point remainder
function becomes unity. This choice differs from the normalization
used in~\cite{Bartels:2014jya} by a factor of $2i$.}
\begin{multline}
f_{k}(\varepsilon_5,\dots,\varepsilon_{k+4};w_5,\dots,w_{k+4})
\equiv\includegraphicsbox{Fignptcut}
=\\
i\,g\sum_{n_5,\dots,n_{k+4}}\int\dd\nu_5\dots\dd\nu_{k+4}
\,\Phi_{\mathrm{L},5}
\,\varepsilon_5^{gE_{n_5,\nu_5}}
\,C_5
\,\varepsilon_6^{gE_{n_6,\nu_6}}
\,C_6
\dots
\,C_{k+3}
\,\varepsilon_{k+4}^{gE_{n_{k+4},\nu_{k+4}}}
\,\Phi_{\mathrm{R},k+4}
\,.
\label{eq:cutintegral}
\end{multline}
One can see that the exponentials of
kinematic variables in the impact factors and emission blocks indeed
combine into Fourier--Mellin integral transformation kernels
\begin{equation}
w_k^{i\nu_k+n_k/2}\,\bar w_k^{i\nu_k-n_k/2}
=
\rho_k^{2i\nu_k}\,e^{in_k\varphi_k}
\qquad
\text{for}
\qquad
w_k=\rho_ke^{i\varphi_k}
\,.
\end{equation}
%

\paragraph{Perturbative Expansion.}

The expression~\eqref{eq:cutintegral} is valid to all orders in the
coupling $g$, where all coupling dependence is contained in
the impact factors $\Phi\subrm{L,R}$,
the emission blocks $C_k$, and the BFKL eigenvalues $E_{n_k,\nu_k}$.
Upon a perturbative expansion, the BFKL
Green's functions~\eqref{eq:bfklgreen} expand in powers of
$g$ and of $\log(\varepsilon_k)$; the latter are the large
logarithms that are characteristic of the multi-Regge limit. Including
subleading terms of the BFKL eigenvalues, impact factors, and emission
blocks, the cut contribution~\eqref{eq:cutintegral} at each order
$g^\ell$ in the coupling constant becomes a polynomial of degree $(\ell-1)$ in
the large logarithms $\log\varepsilon_k$. Retaining only the leading
terms in large logarithms amounts to the leading logarithmic
approximation (LLA), the first subleading terms constitute the
next-to-leading logarithmic approximation (NLLA), and so on. At order
$g^\ell$, there are LLA terms of order $\log(\varepsilon_k)^{\ell-1}$
all the way to N$^{\ell-1}$LLA terms of order
$\log(\varepsilon_k)^{0}$. At a given loop order, the coefficient of
each monomial in $\log(\varepsilon_k)$ is a function of the kinematics
that exclusively depends on the complex anharmonic ratios
$w_k$~\eqref{eq:wj}.

\section{Symbols and Regions}
\label{sec:symbols}

\paragraph{Transcendentality and Symbols.}

Scattering amplitudes in planar $\superN=4$ super Yang--Mills theory
display the property of uniform (or maximal) transcendentality, which
means that every term in the $\ell$-loop amplitude has the same
transcendentality (or transcendental weight) $2\ell$.
This concept relies on the assumption that the amplitude can be
expanded in products of multiple polylogarithms (iterated integrals
over $\dlog$ integrands, MPLs for short)~\cite{Goncharov:2001iea},
$\pi$, and zeta values.%
\footnote{It is expected that this class of functions is not
sufficient to describe all amplitudes to all orders in general
kinematics. For example, elliptic integrals appear in the ten-point
N$^3$MHV amplitude~\cite{CaronHuot:2012ab}. However, based on the
singularity structure of the integrand, it is safe to assume
that all MHV amplitudes can be expressed as rational polynomials of
multiple polylogarithms and zeta
values~\cite{ArkaniHamed:2010kv,ArkaniHamed:2010gh}, and multi-Regge
limit MHV amplitudes inherit this property.}
Every $m$-fold iterated integral is assigned transcendentality
$m$. Zeta values can be defined as MPLs evaluated on certain values,
and they inherit the transcendentality of their parent functions.
For example, the polylogarithms $\Li_m(x)$ as well as the zeta values
$\zeta_m$ have transcendentality $m$, and $\pi$ has transcendental
weight $1$.
Under multiplication, transcendentality behaves additively.

Multiple polylogarithms obey many functional identities, which makes
them unwieldy, especially in expressions with many terms.
All such functional relations trivialize when one projects all MPLs to
their \emph{symbols}~\cite{Goncharov.A.B.:2009tja}.%
\footnote{For reviews, see~\cite{Duhr:2011zq,Duhr:2014woa}.}
The latter discard all information contained in the choice of
integration base point. In particular, the symbols are agnostic of all
ambiguities lying in the choice of functional branch. Since all branch
ambiguities of MPLs have subleading \emph{functional}
transcendentality (transcendentality of functional origin, as opposed
to numerical transcendentality), one typically discards
\emph{all} terms of subleading functional transcendentality when
mapping an expression to its symbol.

When projecting the amplitude to its symbol, the
expression~\eqref{eq:allcuts} simplifies considerably: The reduced
pole terms consist of trigonometric functions whose arguments include
factors of $\pi$~\cite{Bartels:2013jna,Bartels:2014jya}, hence their
perturbative expansion
contains extra powers of $\pi$, which implies that they carry
subleading functional transcendental weight; they therefore get
discarded. Cut terms that involve more than two Reggeons stem from
double (or higher) discontinuities, hence they also have subleading
transcendentality and get projected out. Terms with multiple
disconnected multi-Reggeon states (such as the last term
in~\eqref{eq:allcuts}) are products of lower-loop cut terms, hence
also these have subleading transcendental weight and get discarded. On
the left hand side of the equation, the factor $\exp\brk{i\delta_n^\rho}$ can be
truncated to $1$, since all higher terms again include additional
factors of $\pi$. In summary, at the level of the symbol, the
remainder function is a linear combination of two-Reggeon cut terms:
\begin{equation}
R_n^\rho
\simeq
\sum_jc_{n,1,j}^\rho\includegraphicsbox{Fignpt2r}
+\sum_jc_{n,2,j}^\rho\includegraphicsbox{Fignpt22r}
+\dots
\label{eq:Rcuts}
\end{equation}
Here, ``$\simeq$'' denotes equality at the symbol level.%
\footnote{Strictly speaking, the symbol vanishes, since the right-hand
side contains an overall factor of $2\pi i$. What is meant by
``$\simeq$'' is that the symbols on both sides agree after pulling out
the overall $2\pi i$ factor.}
Moreover, here and in the following, the remaining (black) cut pieces
are understood to be one-loop subtracted, as the one-loop part is
(by definition) contained in the BDS factor that has been divided out.
The dots stand for further two-Reggeon cut terms that span any number
of adjacent emitted gluons.

\paragraph{Symbols and Regions.}

At the symbol level,
the discontinuity of an iterated integral along a closed continuation path only depends
on the overall winding numbers of the path around the singular points
of the integrand.
From this property alone, it follows~\cite{Bargheer:2015djt} that the
symbols $\smbop{\cdot}$ of the multi-Regge-limit remainder function in the various
kinematic regions obey the
relations
\begin{equation}
\smbop{R_n^I}
=
\sum_{\brc{k,l}\subset I}
\smbop{R_n^{\brc{k,l}}}\, .
\label{eq:rel1}
\end{equation}
and
\begin{equation}
\smbop{R_n^{\brc{k,l}}}
=
\smbop{R_n^{[k,l]}}
-\smbop{R_n^{[k,l-1]}}
-\smbop{R_n^{[k+1,l]}}
+\smbop{R_n^{[k+1,l-1]}}
\,.
\label{eq:rel2}
\end{equation}
These relations hold independently of the loop order.
The first relation states that the symbol in any region
$I\subset\brc{4,\dots,n-1}$ is a sum of symbols in regions
$\brc{k,l}$ where only two momenta $\bfm{p}_k$ and $\bfm{p}_l$ are
flipped.
The second relation in turn expresses the symbol in those two-flip
regions as a linear combination of symbols in regions
where all
flipped momenta $k,\dots,l$ are adjacent, labeled by $\sbrk{k,l}$.
It is therefore sufficient to consider the symbol in those
all-adjacent regions.

Note that, since the cut terms can be assumed to be functionally
independent, the relations~\eqref{eq:rel1,eq:rel2} among symbols imply
identical relations for the cut prefactors
$c_{n,b,j}^\rho$ in the various regions:
\begin{equation}
c_{n,b,j}^I
=
\sum_{\brc{k,l}\subset I}
c_{n,b,j}^{\brc{k,l}}
\,,\qquad
c_{n,b,j}^{\brc{k,l}}
=
 c_{n,b,j}^{[k,l]}
-c_{n,b,j}^{[k,l-1]}
-c_{n,b,j}^{[k+1,l]}
+c_{n,b,j}^{[k+1,l-1]}
\,.
\label{eq:rel12c}
\end{equation}
It is not difficult to see that these relations are consistent with
the Mandelstam criterion~\eqref{eq:Mandelstamcrit} described above. They completely determine
the coefficients of all two-Reggeon cut contributions of the type shown
in~\eqref{eq:Rcuts} to the $n$-point remainder function in any
kinematic region $\rho$ in terms of the coefficients
$c_{n,b,j}^{\sbrk{k,l}}$ of these cut terms in the all-adjacent
regions $\rho=\sbrk{k,l}$.

In fact, the Mandelstam criterion~\eqref{eq:Mandelstamcrit}
implies that there is only a single two-Reggeon cut contribution to the
$n$-point multi-Regge limit remainder function in any
all-adjacent region $\sbrk{k,l}$, namely
\begin{equation}
R_{n,\mathrm{cut}}^{\sbrk{k,l}}
=c_{n,l-k,k}^{\sbrk{k,l}}
\includegraphicsbox{Fignpt2anyr}
\,,
\label{eq:ncut}
\end{equation}
where the subscript ``cut'' indicates that the Regge pole terms are
not included, and
the dots stand for the omission of $(l-k-2)$ emission
blocks. In other words,
\begin{equation}
c_{n,b,j}^{\sbrk{k,l}}=0
\qquad
\text{unless}
\qquad
j=k\,,
\quad
\text{and}
\quad
b=l-k
\,.
\end{equation}
In particular, the cut terms in all such regions equal (up to variable substitution and
the prefactors)
the cut terms of the
$(l-k+5)$-point remainder function in the region where \emph{all} intermediate momenta
are flipped:
\begin{equation}
R_{n,\mathrm{cut}}^{\sbrk{k,l}}(\varepsilon_{k+1},\dots,\varepsilon_{l};w_{k+1},\dots,w_{l})
=
\frac{c_{n,l-k,k}^{\sbrk{k,l}}}{c_{n',n'-4,4}^{\sbrk{4,n'-1}}}
R_{n',\mathrm{cut}}^{\sbrk{4,n'-1}}(\varepsilon_{k+1},\dots,\varepsilon_{l};w_{k+1},\dots,w_{l})
\,,
\label{eq:pointred}
\end{equation}
with $n'=l-k+5$.
Since the symbol of the remainder funtion is agnostic of the pole
terms, the equations~\eqref{eq:ncut} and~\eqref{eq:pointred} hold for
the full remainder function at the symbol level.

\section{Two-Loop Expansion}
\label{sec:2loop}

We now want to analyze the two-Reggeon
contribution~\eqref{eq:ncut} for any number of gluons
at the perturbative level. The following deconstruction is not
restricted to symbols, but holds at the level of full
functions.
By definition, all cut diagrams of the
type~\eqref{eq:ncut} are understood to be one-loop subtracted. The simplest case
involves only two emitted gluons. Perturbatively expanding the BFKL Green's function
and the impact factors, this simplest diagram consists of
three terms at the two-loop level:
\begin{equation}
f_1(\varepsilon_5;w_5)
=\includegraphicsbox{Fig6ptcut}
=\includegraphicsbox{Fig6ptp010}
+\includegraphicsbox{Fig6ptp100}
+\includegraphicsbox{Fig6ptp001}
+\order{g^3}
\,.
\label{eq:2loop6pt}
\end{equation}
Here, a naked line for the
impact factor stands for its leading contribution~\eqref{eq:iflead},
whereas additional dots denote loop corrections.
A vertical line in the $t$-channel two-Reggeon state stands for
the one-loop (order $g^1$) piece of the BFKL Green's
function~\eqref{eq:bfklgreen},
\begin{equation}
G_2(n_k,\nu_k)=\varepsilon_k^{gE_{n_k,\nu_k}}
=
1+g\,E_{n_k,\nu_k}^{(0)}\log(\varepsilon_k)
+\order{g^2}
\,,
\label{eq:gfexpansion1}
\end{equation}
where
\begin{equation}
E_{n_k,\nu_k}
=
\sum_{\ell=0}^\infty
g^{\ell}E_{n_k,\nu_k}^{(\ell)}
\end{equation}
is the expansion of the BFKL eigenvalue. Due to the factor
$\log(\varepsilon_k)$ in the
one-loop Green's function,
the first term in~\eqref{eq:2loop6pt} provides the leading logarithmic approximation
(LLA) at this two-loop order.
The subleading NLLA contribution consists of the second and third
diagrams, which have
no line insertions, and stem from the trivial piece
$G_2(g=0)=1$ of the Green's function.

Turning to the longer two-Reggeon cut that appears in the $(\m\m\m)$ region of the
seven-point remainder function, the two-loop
expansion yields five terms,
\begin{multline}
f_2(\varepsilon_5,\varepsilon_6;w_5,w_6)
=\includegraphicsbox{Fig7ptcut}
=\includegraphicsbox{Fig7ptp01000}
+\includegraphicsbox{Fig7ptp00010}
\\
+\includegraphicsbox{Fig7ptp10000}
+\includegraphicsbox{Fig7ptp00100}
+\includegraphicsbox{Fig7ptp00001}
+\order{g^3}
\,.
\label{eq:7point2loop}
\end{multline}
Here, the emission block makes its first appearance. A plain dotted
line stands for the leading-order emission block~\eqref{eq:evlead},
and additional dots
again denote loop corrections. The LLA piece now consists of
two terms, where either of the Green's functions in the first or
second $t$-channel have been expanded to one-loop order. Hence the
first term is proportional to $\log(\varepsilon_5)$, while the second
term is proportional to $\log(\varepsilon_6)$.%
\footnote{Without loss of generality, it is assumed that the leftmost
particle at the beginning of the cut is particle $4$.}
The second
line provides the three NLLA terms.

A key fact for the subsequent analysis is the
following observation~\cite{Bartels:2011ge}: Any number of adjacent leading-order
emission blocks, not separated by BFKL eigenvalue insertions, can be
absorbed in a neighboring leading-order impact factor (again not
separated by BFKL eigenvalue insertions). The result is the original
impact factor, whose momentum gets replaced by the sum of combined
momenta. Similarly, any number of adjacent leading-order
emission blocks can be combined into a multi-gluon emission
block, whose functional form is identical to the single-gluon block,
but whose outgoing momentum is replaced by the sum of all combined
momenta.
Diagrammatically, we will denote these identities~as
\begin{equation}
\includegraphicsbox{Figifpvid1}
\equiv
\includegraphicsbox{Figifpvid2}
\,,\qquad
\text{and}
\qquad
\includegraphicsbox{Figpvid1}
\equiv
\includegraphicsbox{Figpvid2}
\,.
\label{eq:ifpvreduction}
\end{equation}
Here, the dots stand for the insertion of any number of leading-order
emission blocks.
The identity for impact factors (left) was demonstrated
in~\cite{Bartels:2011ge}, and the identity for emission blocks (right)
follows straightforwardly. For completeness, the identities are derived
in~\appref{app:reduction-identities}.
Using these identities, one can reduce almost all diagrams
in~\eqref{eq:7point2loop} to six-point diagrams. For example,
\begin{equation}
\includegraphicsbox{Fig7ptp01000}
=
\includegraphicsbox{Fig6ptp010l12}
\,,\qquad
\includegraphicsbox{Fig7ptp00010}
=
\includegraphicsbox{Fig6ptp010l21}
\,.
\label{eq:reduction}
\end{equation}
Each term in the two-loop expression~\eqref{eq:7point2loop}
\emph{a priori} depends on both complex anharmonic ratios
$w_5$ and $w_6$~\eqref{eq:wj}. But due to
the identity~\eqref{eq:reduction}, it is clear that all dependence of the first term
in~\eqref{eq:7point2loop} on $w_5$ and $w_6$ factors into a dependence
on the single complex ratio
\begin{equation}
v_{5,6;5}\equiv\frac{p_4q_7}{q_4(p_5+p_6)}=\frac{w_5}{\bigbrk{1+\frac{1}{w_6}}}\,.
\label{eq:v4}
\end{equation}
Similarly, the second term in~\eqref{eq:7point2loop} only depends
on the single complex ratio
\begin{equation}
v_{5,6;6}\equiv\frac{(p_4+p_5)q_7}{q_4p_6}=(1+w_5)w_6\,.
\label{eq:v5}
\end{equation}
Restricting to the LLA (the first line in~\eqref{eq:7point2loop}), and
using the identities~\eqref{eq:reduction}, the
three-particle cut
therefore reduces to a sum of two copies of the
two-particle cut,
\begin{equation}
f_{2,(2)}\suprm{LLA}(\varepsilon_5,\varepsilon_6;w_5,w_6)
=
 f_{1,(2)}\suprm{LLA}(\varepsilon_5;v_{5,6;5})
+f_{1,(2)}\suprm{LLA}(\varepsilon_6;v_{5,6;6})
\,.
\end{equation}
Promoting this equation to the full two-loop
cut contribution
(including the NLLA piece) requires adding an extra NLLA term to the
equation:
\begin{equation}
f_{2,(2)}(\varepsilon_5,\varepsilon_6;w_5,w_6)
=
 f_{1,(2)}(\varepsilon_5;v_{5,6;5})
+f_{1,(2)}(\varepsilon_6;v_{5,6;6})
+g_2(v_{5,6;5},v_{5,6;6})\,,
\end{equation}
where
\begin{equation}
g_2(v_{5,6;5},v_{5,6;6})
=\includegraphicsbox{Fig7ptp00100}
-\includegraphicsbox{Fig6ptp100l21}
-\includegraphicsbox{Fig6ptp001l12}
\label{eq:gdiag}
\end{equation}
is a finite function of $v_{5,6;5}$ and $v_{5,6;6}$ (or of $w_4$ and
$w_5$ via the relations~\eqref{eq:v4,eq:v5}). Here, the last two terms
appear in the two $f_{1,(2)}$ terms but not in $f_{2,(2)}$ and thus
need to be subtracted. They are,
by analogy with~\eqref{eq:ifpvreduction}, defined by evaluating the
one-loop impact factors on the sums of momenta $p_4+p_5$ and
$p_5+p_6$, respectively.

This two-loop analysis straightforwardly generalizes to the cut
contribution for any number of particles.
Using the identities~\eqref{eq:ifpvreduction}, the LLA part of the
general cut at two loops can be written as
\begin{equation}
f_{k,(2)}\suprm{LLA}(\varepsilon_5,\dots,\varepsilon_{4+k};w_5,\dots,w_{4+k})
=
\sum_{j=5}^{4+k}
\,\includegraphicsbox{Fig6ptp010lgen}
=
\sum_{j=5}^{4+k}
f_{1,(2)}\suprm{LLA}(\varepsilon_{j};v_{5,4+k;j})
\,.
\label{eq:nptLLA}
\end{equation}
Here, the variables
\begin{equation}
v_{k,l;j}
\equiv
\frac{q_{k-1}-q_{j}}{q_{k-1}}\frac{q_{l+1}}{q_{j}-q_{l+1}}
=
\frac{%
\brk{1+\brk{1+\brk{\dots\brk{1+w_{k}}w_{k+1}}\dots}w_{j-1}}w_j%
}{%
1+\lrbrk{1+\lrbrk{\dots\lrbrk{1+\frac{1}{w_{l}}}\frac{1}{w_{l-1}}}\dots}\frac{1}{w_{j+1}}%
}
\label{eq:vmap}
\end{equation}
for $j=k,\dots,l$
are anharmonic ratios that generalize~\eqref{eq:v4,eq:v5}; they are
obtained by
grouping the adjacent momenta $p_{k-1}+\dots+p_{j-1}=q_{k-1}-q_j$ and
$p_{j}+\dots+p_l=q_j-q_{l+1}$.
The inversion of this formula is
\begin{equation}
w_j=\frac{(v_{k,l;j-1}-v_{k,l;j})(1+v_{k,l,j+1})}{(1+v_{k,l;j-1})(v_{k,l;j}-v_{k,l;j+1})}
\,,
\end{equation}
assuming the boundary conditions $v_{k,l;k-1}=0$ and $v_{k,l;l+1}=\infty$.
Including the NLLA terms of $f_{k,(2)}$ and $f_{1,(2)}$ on both
sides of equation~\eqref{eq:nptLLA}, and again applying the
reduction identities~\eqref{eq:ifpvreduction}, one can see
that all subleading terms combine into a sum of seven-point NLLA
pieces $g_2$~\eqref{eq:gdiag}, evaluated with different complex
ratios:
\begin{equation}
f_{k,(2)}(\varepsilon_5,\dots,\varepsilon_{4+k};w_5,\dots,w_{4+k})
=
\sum_{j=5}^{4+k}
f_{1,(2)}(\varepsilon_j;v_j)
+
\sum_{j=5}^{3+k}
g_2(v_j,v_{j+1})
\,,
\qquad
v_j\equiv v_{5,4+k;j}
\,.
\label{eq:fk2c}
\end{equation}
This concludes the two-loop analysis of the general two-Reggeon
cut~\eqref{eq:ncut}. For any number of emitted particles, the latter can be
deconstructed into a sum of two building blocks, one of them being the
simplest two-particle cut $f_{1}$, the other being the NLLA
remainder $g_2$ of the three-particle cut $f_{2}$.

Using equation~\eqref{eq:ncut}, the result~\eqref{eq:fk2c} directly
implies an analogous relation for the cut piece of the two-loop remainder function
in the region $(\m\m\dots\m)$ where all momenta have been flipped,
\begin{equation}
R_{n,(2),\mathrm{cut}}^{\mathrm{MRL},(\m\m\dots\m)}(\varepsilon_5,\dots,\varepsilon_{n-1};w_5,\dots,w_{n-1})
=
\frac{c_n}{c_6}
\sum_{j=5}^{n-1}
R_{6,(2),\mathrm{cut}}^{\mathrm{MRL},(\m\m)}(\varepsilon_j;v_j)
+
c_n
\sum_{j=5}^{n-2}
g_2(v_j,v_{j+1})
\,,
\label{eq:Rn2c}
\end{equation}
with the abbreviations $c_n\equiv c_{n,n-5,4}^{\sbrk{4,n-1}}$ and
$v_j\equiv v_{5,n-1;j}$. With the help of~\eqref{eq:pointred}, very
similar relations hold for the remainder function \emph{symbol}
in any region $\sbrk{k,l}$ where any number
of adjacent momenta have been flipped.

\paragraph{Relation to Previous Work.}

At leading logarithmic order, the relation~\eqref{eq:fk2c} together
with the variable map~\eqref{eq:vmap} has been
obtained before~\cite{Bartels:2011ge}. Here, we have generalized it to
the full two-loop level, including the NLLA terms. In fact,
an explicit study~\cite{Bargheer:2015djt} of the known two-loop
symbol~\cite{CaronHuot:2011ky} has lead to the slightly stronger
observation
\begin{equation}
R_{n,(2),\mathrm{cut}}^{\mathrm{MRL},(\m\m\dots\m)}(\varepsilon_5,\dots,\varepsilon_{n-1};w_5,\dots,w_{n-1})
\simeq
\sum_{j=5}^{n-1}
R_{6,(2),\mathrm{cut}}^{\mathrm{MRL},(\m\m)}(\varepsilon_j;v_j)
+
c_7
\sum_{j=5}^{n-2}
g_2(v_j,v_{j+1})
\,.
\label{eq:Rn2}
\end{equation}
Also this result had been obtained previously at leading logarithmic
order~\cite{Prygarin:2011gd}.
Comparing~\eqref{eq:Rn2c} with~\eqref{eq:Rn2}, one finds that the
coefficients of all simple two-Reggeon cut contributions must be identical,%
\footnote{Since~\cite{Bargheer:2015djt} analyzed the two-loop symbol
for up to ten points, the equality has only been rigorously
established for $n\leq10$.}
\begin{equation}
c_{n,l-k,k}^{\sbrk{k,l}}=c_{6,1,4}^{\sbrk{4,5}}=1
\,,\qquad
n\geq7
\,,\quad
4\leq k<l<n
\,.
\label{eq:cvals}
\end{equation}
Here, the second equality follows from the deliberate choice of
normalization~\eqref{eq:cutintegral} for the cut integral.

\section{Three-Loop Expansion}
\label{sec:3loop}

We are now in a position to extend the previous analysis to the
three-loop order.
At three loops, the simplest cut contribution $f_1$
expands to
\begin{multline}
f_{1,(3)}=
 \includegraphicsbox{Fig6ptp020}
+\includegraphicsbox{Fig6ptp110}
+\includegraphicsbox{Fig6ptp01l0}
+\includegraphicsbox{Fig6ptp011}
\\
+\includegraphicsbox{Fig6ptp200}
+\includegraphicsbox{Fig6ptp101}
+\includegraphicsbox{Fig6ptp002}
\,.
\label{eq:f13}
\end{multline}
Compared to the two-loop case, there are a few new ingredients at
three loops: Two
line insertions in the two-Reggeon state (as in the first term) stand for terms where the
BFKL Green's function~\eqref{eq:bfklgreen} has been expanded to
second order in the coupling $g$, while the BFKL eigenvalue $E_{n,\nu}$ has
been kept at leading order. A line insertion dressed with a dot stands
for one power of the one-loop correction to the eigenvalue
$E_{n,\nu}$. Each line (leading order or loop corrected) comes with
one power of the respective large logarithm $\log(\varepsilon_k)$. In
other words, expanding
\begin{multline}
G_2(n_k,\nu_k)=\varepsilon_k^{gE_{n_k,\nu_k}}
=\\
1+g\,E_{n_k,\nu_k}^{(0)}\log(\varepsilon_k)
+\sfrac{1}{2} g^2\bigbrk{E_{n_k,\nu_k}^{(0)}\log(\varepsilon_k)}^2
+g^2E_{n_k,\nu_k}^{(1)}\log(\varepsilon_k)
+\order{g^3}
\,,
\label{eq:gfexpansion}
\end{multline}
where
$E_{n_k,\nu_k}^{(\ell)}$
is the $\ell$-loop BFKL eigenvalue, the third term
in~\eqref{eq:gfexpansion} produces the first term
in~\eqref{eq:f13}, whereas the fourth term in~\eqref{eq:gfexpansion}
produces the third term in~\eqref{eq:f13}.
The first term in~\eqref{eq:f13} constitutes the LLA part, the next
three terms provide the NLLA contribution, and the three terms on the
second line form the NNLLA piece.

Passing now to the longer cut $f_2$, one finds the following terms at
three loops and leading logarithmic order:
\begin{multline}
f_{2,(3)}
=\includegraphicsbox{Fig7ptp02000}
+\includegraphicsbox{Fig7ptp01010}
+\includegraphicsbox{Fig7ptp00020}
+\order{\mathrm{NLLA}}
\\
=\includegraphicsbox{Fig6ptp020l12}
+\includegraphicsbox{Fig7ptp01010}
+\includegraphicsbox{Fig6ptp020l21}
+\order{\mathrm{NLLA}}
\,.
\end{multline}
As shown in the second line, two of the LLA diagrams can again be reduced to
six-point diagrams, using~\eqref{eq:ifpvreduction}. But, unlike in the two-loop case, one LLA diagram
remains that cannot be reproduced by six-point data.
Removing the six-point pieces by subtracting two instances of
$f_{1,(3)}$ functions~\eqref{eq:f13}, one finds the remainder
(without loss of generality, the emitted gluons are labeled by
$\brc{4,5,6}$)
\begin{multline}
g_3(\varepsilon_5,\varepsilon_6;w_5,w_6)
\equiv
 f_{2,(3)}(\varepsilon_5,\varepsilon_6;w_5,w_6)
-f_{1,(3)}(\varepsilon_5;v_{5,6;5})
-f_{1,(3)}(\varepsilon_6;v_{5,6;6})
\\
=\includegraphicsbox{Fig7ptp01010}
+\includegraphicsbox{Fig7ptp01100}
+\includegraphicsbox{Fig7ptp00110}
+\includegraphicsbox{Fig7ptp01001}
+\includegraphicsbox{Fig7ptp10010}
\\
-\includegraphicsbox{Fig6ptp110l21}
-\includegraphicsbox{Fig6ptp011l12}
+\includegraphicsbox{Fig7ptp10001}
+\includegraphicsbox{Fig7ptp10100}
+\includegraphicsbox{Fig7ptp00101}
\\
+\includegraphicsbox{Fig7ptp00200}
-\includegraphicsbox{Fig6ptp101l21}
-\includegraphicsbox{Fig6ptp101l12}
-\includegraphicsbox{Fig6ptp200l21}
-\includegraphicsbox{Fig6ptp002l12}
\,.
\label{eq:g3def}
\end{multline}
Note that all terms involving the next-to-leading-order BFKL
eigenvalue are captured by the short cut terms $f_{1(3)}$.
It is now straightforward to see that the general $k$-point cut diagram
$f_{k,(3)}$, to leading
logarithmic order,
becomes a sum of six-point functions $f_{1,(3)}$ and
seven-point functions $g_3$:%
\footnote{Note that, contrary to the two-loop case~\eqref{eq:gdiag},
the three-loop building block $g_3$ is defined in terms of the
original cross ratios $w_5$, $w_6$ rather than the combinations
$v_{5,6;5}$, $v_{5,6;6}$.}
\begin{multline}
f_{k,(3)}\suprm{LLA}(\varepsilon_5,\dots,\varepsilon_{4+k};w_5,\dots,w_{4+k})
\\
=
\sum_{j=5}^{4+k}
f_{1,(3)}\suprm{LLA}(\varepsilon_j;v_{5,4+k;j})
+
\sum_{i=5}^{3+k}
\sum_{j=i+1}^{4+k}
g_3\suprm{LLA}(\varepsilon_i,\varepsilon_j;v_{5,j-1;i},v_{i+1,4+k;j})
\,.
\end{multline}
Including all NLLA and NNLLA diagrams in the functions $f_{k,(3)}$,
$f_{1,(3)}$, and $g_3$ on both sides of the above equation, and judiciously
organizing all terms, one finds that
the subleading contributions can be combined into two
further NLLA building blocks $g_{\mathrm{L}}$, $g_{\mathrm{R}}$, and
one further NNLLA building block $h$. The full three-loop cut function
$f_{k,(3)}$ can be written as
\begin{multline}
\!f_{k,(3)}(\varepsilon_5,\dots,\varepsilon_{k+4};w_5,\dots,w_{k+4})
=
\sum_{j=5}^{k+4}
f_{1,(3)}(\varepsilon_j;v_{5,4+k;j})
+
\sum_{i=5}^{k+3}
\sum_{j=i+1}^{k+4}
g_3(\varepsilon_i,\varepsilon_j;v_{5,j-1;i},v_{i+1,k+4;j})
\\
+
\sum_{i=5}^{k+2}
\sum_{j=i+1}^{k+3}
g_{\mathrm{L}}(\varepsilon_i;v_{5,j-1;i},v_{i+1,j;j},v_{j+1,k+4;j+1})
+
\sum_{i=5}^{k+2}
\sum_{j=i+2}^{k+4}
g_{\mathrm{R}}(\varepsilon_{j};v_{5,i;i},v_{i+1,j-1;i+1},v_{i+2,k+4;j})
\\
+
\sum_{i=5}^{k+1}
\sum_{j=i+2}^{k+3}
h(v_{5,i;i},v_{i+1,j-1;i+1},v_{i+2,j;j},v_{j+1,k+4;j+1})
\,.
\label{eq:fk3}
\end{multline}
The NLLA building block $g\subrm{L}$
depends on four intermediate momenta. It takes the form
\begin{equation}
g_{\mathrm{L}}(\varepsilon_5;w_5,w_6,w_7)
=\includegraphicsbox{Fig8ptp0100100}
-\includegraphicsbox{Fig7ptp01001l112}
-\includegraphicsbox{Fig7ptp01100l121}
+\includegraphicsbox{Fig6ptp011l13}
\,.
\label{eq:gL}
\end{equation}
In the third term of~\eqref{eq:fk3}, this function
gets summed over partitions of the sequence of momenta
$\brk{p_4,\dots,p_{k+4}}$ into subsequences
\begin{equation}
\brk{p_4,\dots,p_{i-1}}\,,
\quad
\brk{p_{i},\dots,p_{j-1}}\,,
\quad
\brk{p_{j}}\,,
\quad
\text{and}
\quad
\brk{p_{j+1},\dots,p_{k+4}}
\,.
\end{equation}
The building block $g\subrm{R}$ is a mirror of $g\subrm{L}$:
\begin{equation}
g_{\mathrm{R}}(\varepsilon_7;w_5,w_6,w_7)
=\includegraphicsbox{Fig8ptp0010010}
-\includegraphicsbox{Fig7ptp10010l211}
-\includegraphicsbox{Fig7ptp00110l121}
+\includegraphicsbox{Fig6ptp110l31}
\,,
\label{eq:gR}
\end{equation}
and in the fourth term of~\eqref{eq:fk3}, it gets summed over the partitions
\begin{equation}
\brk{p_4,\dots,p_{i-1}}\,,
\quad
\brk{p_{i}}\,,
\quad
\brk{p_{i+1},\dots,p_{j-1}}\,,
\quad
\text{and}
\quad
\brk{p_{j},\dots,p_{k+4}}
\,.
\end{equation}
Finally, the N$^2$LLA
building block $h$ reads:
\begin{multline}
h(w_5,w_6,w_7,w_8)
=\includegraphicsbox{Fig9ptp001000100}
-\includegraphicsbox{Fig8ptp1000100l2111}
-\includegraphicsbox{Fig8ptp0010001l1112}
\\
+\includegraphicsbox{Fig8ptp1000100l1211}
+\includegraphicsbox{Fig8ptp0010001l1121}
-\includegraphicsbox{Fig7ptp10001l221}
-\includegraphicsbox{Fig7ptp10001l122}
\\
-\includegraphicsbox{Fig7ptp10100l131}
-\includegraphicsbox{Fig7ptp00101l131}
-\includegraphicsbox{Fig7ptp00200l131}
+\includegraphicsbox{Fig7ptp10001l212}
\\
+\includegraphicsbox{Fig6ptp200l41}
+\includegraphicsbox{Fig6ptp101l41}
+\includegraphicsbox{Fig6ptp101l14}
+\includegraphicsbox{Fig6ptp002l14}
\,.
\label{eq:h}
\end{multline}
The last term in~\eqref{eq:fk3} sums this function
over partitions of the intermediate momenta
into subsequences
\begin{equation}
\brk{p_4,\dots,p_{i-1}}\,,
\quad
\brk{p_{i}}\,,
\quad
\brk{p_{i+1},\dots,p_{j-1}}\,,
\quad
\brk{p_{j}}\,,
\quad
\text{and}
\quad
\brk{p_{j+1},\dots,p_{k+4}}
\,.
\end{equation}
For the case $k=3$, which is relevant for the eight-point remainder
function in the $(\m\m\m\m)$ region, the last sum in~\eqref{eq:fk3} has to be
replaced by the single term $\tilde h(w_5,w_6,w_7)$, where $\tilde h$
is obtained from $h$~\eqref{eq:h} by removing the middle particle
(and the associated LO emission block, if applicable).

Using~\eqref{eq:ncut}, the deconstruction~\eqref{eq:fk3} implies an analogous
relation for the three-loop remainder function in the region
$\rho=\sbrk{4,n-1}=(\m\m\dots\m)$ where all
intermediate momenta have been flipped:
\begin{multline}
R_{n,(3),\mathrm{cut}}^{(\m\m\dots\m)}(\varepsilon_4,\dots,\varepsilon_{n-2};w_4,\dots,w_{n-2})
=
\sum_{j=5}^{n-1}
R_{6,(3),\mathrm{cut}}^{(\m\m)}(\varepsilon_j;v_j)
+
\sum_{\substack{i,j=5\\i<j}}^{n-1}
g_3(\varepsilon_i,\varepsilon_j;v_{5,j-1;i},v_{i+1,n-1;j})
\\
+
\sum_{i=5}^{n-3}
\sum_{j=i+1}^{n-2}
g_{\mathrm{L}}(\varepsilon_i;v_{5,j-1;i},v_{i+1,j;j},v_{j+1,n-1;j+1})
+
\sum_{i=5}^{n-3}
\sum_{j=i+2}^{n-1}
g_{\mathrm{R}}(\varepsilon_{j};v_{5,i;i},v_{i+1,j-1;i+1},v_{i+2,n-1;j})
\\
+
\sum_{i=5}^{n-4}
\sum_{j=i+2}^{n-2}
h(v_{5,i;i},v_{i+1,j-1;i+1},v_{i+2,j;j},v_{j+1,k+4;j+1})
\,.
\label{eq:3loopallterms}
\end{multline}
Here, the identities~\eqref{eq:cvals} among the cut coefficients have already been taken into account.
Via~\eqref{eq:pointred}, equivalent relations hold for the
remainder function in all regions $\rho=\sbrk{k,l}$ where any number
of adjacent momenta $\brc{k,\dots,l}$
has been flipped. In more general regions, the remainder function
receives contributions from further cut terms (of the type shown in
the last line of~\eqref{eq:allcuts}). Passing to the remainder
function \emph{symbol}, these further cut terms drop out (due to their
subleading functional transcendentality). Thus,
by~\eqref{eq:rel1,eq:rel2}, the
deconstruction~\eqref{eq:3loopallterms} implies a decomposition of the
remainder function symbol in \emph{any} kinematic region in terms of the symbols of the five
building blocks $f_{1,(3)}$, $g_3$, $g_L$, $g_R$, and $h$.

\section{Building Blocks}
\label{sec:blocks}

In principle, each term in the perturbative expansion of the Regge
cut diagram~\eqref{eq:ncut} can be computed from the integral
representation~\eqref{eq:cutintegral}, once the expressions for the BFKL
eigenvalue, impact factor, and central emission block are known to
the desired perturbative order. In the previous sections, we have
shown that, by judiciously organizing all terms in the expansion, the
$n$-point two-loop and three-loop cut contributions can be reconstructed
from a few basic building blocks that are functions of the anharmonic
ratios $w_j$. Once these building block functions are known, the Regge
cut contribution to the remainder function can be computed via the
formulas~\eqref{eq:Rn2c,eq:3loopallterms}.

Here, we will content ourselves with treating the building block functions
at the level of the symbol. The symbol of the two-loop NLLA building
block $g_2(v_1,v_2)$ has been obtained~\cite{Bargheer:2015djt} by taking
the multi-Regge limit of the known two-loop remainder function
symbol~\cite{CaronHuot:2011ky} and using the
decomposition~\eqref{eq:Rn2}.

At three loops, both the six-point and
seven-point remainder function symbols are
known~\cite{Dixon:2011pw,Drummond:2014ffa}.
By its definition, this data is sufficient to extract the symbol
of the building block $g_3$~\eqref{eq:g3def}, which contains LLA,
NLLA, and NNLLA parts. Applying in turn the
first line of the three-loop decomposition~\eqref{eq:3loopallterms},
this admits a reconstruction of the $n$-point remainder function
symbol at leading logarithmic order.

We compute the multi-Regge limit symbol of the three-loop remainder
function in the same way as for the two-loop symbol. The procedure is
detailed in~\cite{Bargheer:2015djt}, here we only give a brief
summary: Starting with the known six-point and
seven-point symbols for general kinematics, we
expand all first entries in terms of the cross ratios~\eqref{eq:uj}
via the symbol rule
\begin{equation}
(xy)\otimes(z)
=
(x)\otimes(z)+(y)\otimes(z)
\,.
\label{eq:symbolexpansion}
\end{equation}
Next, we collect all terms with the same cross ratio $U_{k,l}$ in the
first entry, strip off the first entry, and multiply by $2\pi i$. The
result is the symbol of the discontinuity under continuation along the
path $U_{k,l}\to e^{2\pi i}U_{k,l}$. In order to obtain the
multi-Regge limit symbol of each discontinuity, we
express the kinematic invariants in the symbol entries in terms of the
OPE variables
\begin{equation}
\brc{T_j,S_j,F_j}
=
\brc{e^{-\tau_j},e^{\sigma_j},e^{i\phi_j}}
\,,\qquad
j=5,\dots,n-1
\,,
\label{eq:opevars}
\end{equation}
of~\cite{Basso:2013aha},%
\footnote{Compared to~\cite{Basso:2013aha}, we cyclically shift the
momentum twistors, such that $Z_i\suprm{here}=Z_{i+1}\suprm{BSV}$.}
set $S_j=1/(T_jr_j)$, and take the limit $T_j\to0$, keeping
only the leading term in each entry. For the six-point case,
\begin{equation}
r_5^2={w_5\bar w_5}\,,
\qquad
F_5^2={w_5/\bar w_5}\,,
\qquad
T_5^2={\varepsilon_5/r_5}\,,
\end{equation}
whereas for seven points,
\begin{gather}
r_5^2={w_6\bar w_6}\,,
\qquad
F_5^2={w_6/\bar w_6}\,,
\qquad
T_5^2={\varepsilon_6/r_5}\,,
\nn\\
r_6^2=1/{w_5\bar w_5}\,,
\qquad
F_6^2={\bar w_5/w_5}\,,
\qquad
T_6^2={\varepsilon_5/r_6}\,.
\label{eq:frtow7}
\end{gather}
Finally, again expanding
all terms via~\eqref{eq:symbolexpansion}, one can extract all
large logarithms via the shuffle relations
\begin{multline}
\log(\varepsilon_j)\bigbrk{x\otimes y\otimes\dots\otimes z}
=\bigbrk{\varepsilon_j\otimes x\otimes y\otimes\dots\otimes z}
+\bigbrk{x\otimes \varepsilon_j\otimes y\otimes\dots\otimes z}
\\
+\bigbrk{x\otimes y\otimes \varepsilon_j\otimes\dots\otimes z}
+\cdots
+\bigbrk{x\otimes y\otimes\dots\otimes \varepsilon_j\otimes z}
+\bigbrk{x\otimes y\otimes\dots\otimes z\otimes \varepsilon_j}
\,.
\end{multline}
At seven points and three loops, the resulting expression for each
discontinuity is a degree-two polynomial
in $\log(\varepsilon_5)$ and $\log(\varepsilon_6)$, whose coefficients
are symbols with five entries that exclusively depend on $w_5$,
$w_6$, and their complex conjugates.
Starting in the kinematic region $(\p\p\p)$ in which no
intermediate momentum is flipped, each other kinematic region is
associated with specific winding numbers for all cross ratios
$U_{k,l}$. Summing the corresponding discontinuities then yields the
remainder function symbol in the respective kinematic region. In particular,
the region $(\m\m\m)$ that contains the three-particle cut $f_2$, only
the cross-ratio $U_{2,6}$~\eqref{eq:Uij} winds non-trivially.
Applying the change of variables~\eqref{eq:v4,eq:v5} and subtracting
the respective six-point three-loop symbols~\eqref{eq:g3def}, one
finally obtains the symbol of the building block $g_3$.

The NLLA and NNLLA building blocks~\eqref{eq:gL,eq:gR,eq:h} first
appear in the three-loop eight-point and nine-point amplitudes, and
can thus not (yet) be extracted from available perturbative data. In principle
these functions could be computed term by term from the integral
representation~\eqref{eq:cutintegral}. While the BFKL eigenvalue and impact
factor are known explicitly to N$^2$LLA and
N$^3$LLA~\cite{Bartels:2008sc,Fadin:2011we,Lipatov:2010ad,Dixon:2013eka,Dixon:2014voa},
and relating the multi-Regge limit to the Wilson loop
OPE~\cite{Hatsuda:2014oza} led to all-order proposals~\cite{Basso:2014pla},
the missing ingredient is the NLO and
NNLO central emission block~\eqref{eq:evlead}.

In principle, the NLO emission block could be extracted from the
building block $g_2$ by subtracting the two reducible terms and
inverting the Fourier--Mellin transform. This however requires
knowledge of the full function $g_2$, which at present is only known
at leading transcendental weight~\cite{Bargheer:2015djt}.

The attached \mathematica\ file \filename{MRL3LLA.m} contains the
symbols for the building blocks $R_{6,(3)}\supMRL$ and~$g_3$, as
well as a function that reconstructs the three-loop
leading-logarithmic-order remainder function symbol in any kinematic
region from these building blocks.

\paragraph{Note on the Alphabet.}

The three-loop three-particle building block
$g_3(\varepsilon_5,\varepsilon_6;w_5,w_6)$ has the same alphabet
$\aleph$
(letters appearing in the entries of the symbol) as the two-loop
three-particle building block $g_2(w_5,w_6)$:
\begin{equation}
\aleph_w
=
\brc{w_5,1+w_5,w_6,1+w_6,1+w_6+w_5w_6}
\cup\brc{\text{c.c.}}
\,,
\label{eq:alphabetw}
\end{equation}
where ``c.c.'' stands for the complex conjugate set of letters.
Using the expansion~\eqref{eq:3loopallterms}, and expanding all
variables $v_{k,l;j}$ in terms of $w_5,\cdots,w_{n-1}$, the alphabet
(of the terms in the first line) becomes big and complicated for
larger $n$. Had one started with the $n$-point symbol, it would have
been difficult to guess the variable transformation~\eqref{eq:vmap}
that simplifies the alphabet and symbol terms.

Beyond seven points, the full alphabet of the remainder function remains
unknown, even in the multi-Regge limit. At six and seven points, the
alphabet apparently does not change with the loop order, with the full
alphabet already visible at two loops. It appears likely that this
pattern breaks at eight points (beyond the leading logarithmic
approximation), since this is the first instance at
which the three-loop building blocks involve more independent legs
than the two-loop building blocks. It would be interesting to
work out the consequences of the deconstruction~\eqref{eq:3loopallterms}
on the higher-point alphabets in more detail.

\section{The Function \texorpdfstring{$g_3$}{g3}}
\label{sec:g3function}

Clearly, it is desirable to obtain the building blocks of the
multi-Regge limit amplitude at function level. The function
for the three-loop six-point building block $f_{1,(3)}$ has been
derived in~\cite{Dixon:2011pw}. Here, we focus on constructing the
function for the new three-loop seven-point building block $g_3$,
which, together with $f_{1,(3)}$, determines the three-particle cut
$f_{2,(3)}$~\eqref{eq:fk3}. While we will not be able to determine the
function $g_3$ completely, we can severely constrain it using the
knowledge of its symbol as well as further constraints from symmetry
and consistency with the collinear limit.

\paragraph{Structure of the Function $g_3$.}

MHV amplitudes in multi-Regge kinematics are rational polynomials in
multiple polylogarithms, $\pi$, and (multiple) zeta values, where all
occuring monomials have the same (uniform) transcendental
weight~\cite{Kotikov:2004er}.
At loop order $\ell$, the remainder function has
weight $2\ell$. The cut terms that make up the remainder function in the
multi-Regge limit are discontinuities of the full remainder function
and thus carry an overall factor $2\pi i$, which is therefore multiplied by
a function of uniform weight $(2\ell-1)$. Collecting large logarithms,
the function $g_3$ decomposes as follows:%
\footnote{We call the function
$g_3(\varepsilon_5,\varepsilon_6;x,y)\equiv
g_3(\varepsilon_5,\varepsilon_6;w_5(x,y),w_6(x,y))$ by the same name
as $g_3(\varepsilon_5,\varepsilon_6;w_5,w_6)$.}
\begin{equation}
g_3(\varepsilon_5,\varepsilon_6;x,y)
=
2\pi i\sum_{m,n=0}^1
\log(\varepsilon_5)^i\log(\varepsilon_6)^j
\Bigbrk{g_3^{m,n}(x,y)+2\pi i\,h_3^{m,n}(x,y)}
\,.
\label{eq:g3comp}
\end{equation}
At each order in the large logarithms $\log(\varepsilon_j)$, we have
split the function into real parts $g_3^{m,n}$ and imaginary
parts $h_3^{m,n}$, each of which is a fixed-weight combination
of multiple polylogarithms and zeta values with real
coefficients.

\paragraph{Ordinary Multiple Polylogarithms.}

In order to construct the function $g_3$ by matching a general ansatz
to its symbol, we first need an irreducible
basis of multiple polylogarithms of the right class.
Multiple polylogarithms, also called Goncharov
polylogarithms~\cite{Goncharov:2001iea}, can be defined
recursively as iterated integrals
\begin{equation}
\label{eq:Gdef}
G(a_1,\ldots,a_n;z)
\equiv
\begin{cases}
\displaystyle
\frac{1}{n!}\log^n z & \text{if }a_1=\ldots=a_n=0\,,\\[2ex]
\displaystyle
\int_0^z \frac{dt}{t-a_1}G(a_2,\ldots,a_n;t) & \text{otherwise,}
\end{cases}
\end{equation}
with $G(;z)=1$. The sequence of parameters $(a_1,\dots,a_n)$ is called
the weight vector, and the length of the weight vector equals the
transcendental weight (or transcendentality) of the function
$G(a_1,\dots,a_n;z)$. Multiple zeta values are defined in terms of
multiple polylogarithms evaluated at unity, and inherit their
transcendental weight: $\zeta_k$ has weight $k$,
$\zeta_{j,k}$ has weight $j+k$, and so forth. $\pi$ has weight $1$.

As noted in~\cite{Bargheer:2015djt}, using the variables
\begin{equation}
x=-v_6=-(1+w_5)w_6\,,
\qquad
y=-1/v_5=-\frac{1+w_6}{w_5\,w_6}\,,
\label{eq:xy}
\end{equation}
the alphabet~\eqref{eq:alphabetw} of the symbol of $g_3$ becomes
\begin{equation}
\aleph_{xy}
=\brc{x,x-1,y,y-1,xy-1}\cup\brc{\text{c.c.}}
\,.
\end{equation}
Multiple polylogarithms whose symbols draw their entries from this
alphabet belong to the class of two-dimensional harmonic
polylogarithms (2dHPLs)~\cite{Gehrmann:2000zt}. A generating set for
these is given~by%
\footnote{The choice of generating set is not unique. We used a
different basis in~\cite{Bargheer:2015djt}, but found the
choice~\eqref{eq:Gbasis} more suitable for the present analysis.}
\begin{equation}
\bigbrc{G(\vec a,x)\,|\,a_i\in\brc{0,1}}
\cup
\bigbrc{G(\vec a,1/y)\,|\,a_i\in\brc{0,1,x}}
\cup\brc{\text{c.c.}}
\,,
\label{eq:Gbasis}
\end{equation}
where $\brc{\text{c.c.}}$ stands for the complex conjugates of the
previous sets. Multiple polylogarithms satisfy shuffle and stuffle
algebra relations,
hence the above generating set is overcomplete. An irreducible basis
of generators is provided by the subset whose weight vectors form
Lyndon words in the ordered sets of letters $\brc{0,1}$ and
$\brc{0,1,x}$, respectively~\cite{RADFORD1979432}. Including the complex conjugate
generators, the resulting irreducible set consists of $10$, $8$, $20$,
$42$, and $108$ basis functions at weights $1$, $2$, $3$, $4$, and
$5$. Including all possible products of lower-weight functions yields
$10$, $63$, $320$, $1433$, and $5190$ linearly independent terms at
weights $1$, $2$, $3$, $4$, and $5$.

\paragraph{Single-Valuedness.}

Besides the consistency with its known symbol, the function $g_3$ has
to satisfy various constraints. One of them is single-valuedness: Due
to unitarity, a physical amplitude can only have branch points where
one of the cross ratios vanishes (or becomes infinite). Since the
cross ratios are expressed in terms of absolute squares of the complex
variables $w_5$ and $w_6$~\eqref{eq:uofw}, a rotation
$(w_5-z,\bar w_5-\bar z)\to(e^{+2\pi i}(w_5-z),e^{-2\pi i}(\bar
w_5-\bar z))$ around any point $z$ in the complex plane can never let
a cross ratio wind around zero (or infinity). The same is true for
rotations of $w_6$, and therefore also for rotations of $x$ and $y$.
The conclusion is that the coefficient functions~\eqref{eq:g3comp} of $g_3$
must be single-valued functions of the complex variables $x$ and $y$.
This property has been essential for the determination of the
six-point multi-Regge limit to high loop
orders~\cite{Dixon:2012yy,Pennington:2012zj,Dixon:2014voa,Dixon:2014xca}.

One could in principle implement the single-valuedness constraint by
first constructing a function using the basis~\eqref{eq:Gbasis} and
then requiring all monodromies to vanish. However, it turns out that
the single-valuedness constraint can be satisfied directly at the level
of the basis: A suitable basis of single-valued multiple
polylogarithms was recently constructed for any number of points~\cite{DelDuca:2016lad}.%
\footnote{See also~\cite{Broedel:2016kls}.}
We can therefore satisfy the single-valuedness constraint by employing
the single-valued basis, without losing generality.%
\footnote{In the first revision of this paper, I had constructed the
function $g_3$ using the basis~\eqref{eq:Gbasis}, which was the state
of the art at the time the preprint of this paper appeared on the
arXiv. I thank the JHEP referee for requesting a construction based on
the single-valued basis that was published at around the same
time~\cite{DelDuca:2016lad}, and which significantly reduces the
number of free parameters that remain after applying all constraints.}
The single-valued basis can be constructed purely algebraically from
the basis of ordinary multiple polylogarithms~\eqref{eq:Gbasis} using
the Hopf algebra structure that underlies the multiple polylogarithm
algebra~\cite{Goncharov:2001iea}: Each holomorphic element $G$ of the
ordinary basis~\eqref{eq:Gbasis} gets promoted to a single-valued
function $\Gs$ by the single-valued map
\begin{equation}
\bfm{s}:G(\vec a,z)
\mapsto
\Gs(\vec a,z)\equiv
(-1)^{\abs{\vec a}}\mu(\bar S\otimes\id)\Delta G(\vec a,z)
\,,
\label{eq:svmap}
\end{equation}
where $\Delta$ is the coproduct, $\id$ is the identity, $\bar S$
is the complex conjugate of the
antipode map of the Hopf algebra, and $\mu$ denotes the multiplication
operator $\mu(a\otimes b)=a\cdot b$. The details are spelled out in
Section 3.4.3 of~\cite{DelDuca:2016lad}, and we will not reproduce
them here. The antiholomorphic elements
of~\eqref{eq:Gbasis} can equally be promoted to single-valued functions,
which however are not independent from the single-valued functions
generated from the holomorphic elements. A full basis of single-valued
2dHPLs is therefore provided by the single-valued completions of the
holomorphic elements of the ordinary basis~\eqref{eq:Gbasis}. Since
this halves the size of the algebra basis, it significantly reduces
the number of linearly independent elements in a general ansatz at any
fixed weight. For example, while a general (real) weight-five ansatz
constructed from the ordinary basis~\eqref{eq:Gbasis} as well as zeta
values has $6305$ terms (and therefore as many undetermined
coefficients), the corresponding ansatz constructed from the
single-valued basis has only $756$ terms.

To summarize, the single-valued algebra basis that we will employ is
\begin{equation}
\bigbrc{\Gs(\vec a,x)|\vec a\in\lyndon\brc{0,1}}
\cup
\bigbrc{\Gs(\vec a,1/y)|\vec a\in\lyndon\brc{0,1,x}}
\,,
\label{eq:Gsbasis}
\end{equation}
where $\lyndon\brc{0,1}$ and $\lyndon\brc{0,1,x}$ denote the sets of Lyndon
words formed from the ordered sets of letters $\brc{0,1}$ and $\brc{0,1,x}$, respectively.
Every single-valued function $\Gs(\vec a,z)$ is
constructed from the ordinary multiple polylog $G(\vec a,z)$ according
to the algebraic prescription~\eqref{eq:svmap}. In addition, we assume that
$\brc{\zeta_2,\zeta_3,\zeta_{2,3},\zeta_5}$ form the algebraically
independent set of (multiple) zeta values up to weight five.

\paragraph{The Ansatz and Symbol Constraints.}

We start with a general polynomial in single-valued basis
functions~\eqref{eq:Gsbasis} and zeta values, such that all monomials
have identical total weight. Given that the three-loop amplitude in
general kinematics has weight six, and taking into account the overall
factor of $2\pi i$ as well as the large logarithms $\log(\varepsilon_5)$,
$\log(\varepsilon_6)$, one finds that the LLA real part
$g_3^{1,1}$ has to have weight three, the NLLA real parts
$g_3^{1,0}$ and $g_3^{0,1}$ have weight
four, and the NNLLA real part $g_3^{0,0}$ has weight
five. The corresponding imaginary parts have weight one less than the
real parts, due to the extra factor $2\pi i$.
The sizes of the general ansätze for all component functions are
displayed in~\tabref{tab:g3params}.
\begin{table}
\centering
\begin{tabular}{rrrrrrrrr@{\qquad}r}
\toprule
Function & $g_3^{1,1}$ & $h_3^{1,1}$ & $g_3^{1,0}$ & $h_3^{1,0}$ & $g_3^{0,1}$ & $h_3^{0,1}$ & $g_3^{0,0}$ & $h_3^{0,0}$ & total\\
\midrule
General ansatz                & 71 & 20 & 236 & 71 & 236 & 71 & 756 & 236 & 1697 \\
Match to symbol               &  6 & 20 &  25 & 71 &  25 & 71 &  91 & 236 &  545 \\
Parity invariance             &  6 & 16 &  21 & 51 &  21 & 51 &  67 & 151 &  384 \\
Target-projectile symmetry    &  4 & 10 &  21 & 51 &   0 &  0 &   ? &  83 &  236 \\
Vanishing collinear limit I   &  2 &  7 &  15 & 44 &   0 &  0 &  55 &  76 &  199 \\
Consistency with the WLOPE I  &  1 &  5 &  11 & 37 &   0 &  0 &  49 &  71 &  174 \\
Vanishing collinear limit II  &  1 &  5 &  11 & 37 &   0 &  0 &  45 &  65 &  164 \\
Consistency with the WLOPE II &  1 &  5 &  11 & 37 &   0 &  0 &  42 &  59 &  155 \\
\bottomrule
\end{tabular}
\caption{Numbers of free parameters in the components of the function
$g_3$ before imposing constraints, after matching to the known
symbols, and after imposing various constraints. The first two
components constitute the LLA part, the next four functions represent
the NLLA part, and the last two functions form the NNLLA part of the
function $g_3$.}
\label{tab:g3params}
\end{table}
The symbol of the function $g_3$ uniquely fixes all terms in the real
parts $g_3^{m,n}$ with the highest functional weight, that is
all terms that are free of zeta values. We can perform
the match by expanding the single-valued functions $\Gs(\vec a,z)$
into combinations of ordinary multiple polylogarithms
$G(\cdot,\cdot)$, and by applying the symbol map
\begin{multline}
\smbop{G(a_{1},\ldots,a_n;z)}=
\sum_{i=1}^{n} \Bigbrk{
 \smbop{G(a_{1},\ldots,\hat a_i,\ldots,a_n;z)}\otimes (a_i-a_{i-1})\\
-\smbop{G(a_{1},\ldots,\hat a_i,\ldots,a_n;z)}\otimes (a_i-a_{i+1})}\,,
\label{Gsymbol}
\end{multline}
where $a_0=z$, $a_{n+1}=0$, and hatted indices are omitted. The match
to the symbol fixes the majority of terms in the real parts
$g_3^{m,n}$ (see~\tabref{tab:g3params}), but the symbol is
insensitive to all terms of subleading functional weight, including
all terms in the imaginary parts $h_3^{m,n}$.

\paragraph{Parity Invariance and Target-Projectile Symmetry.}

While the terms with subleading functional weight are not seen by the
symbol, they can be constrained by symmetry requirements. Firstly, MHV
amplitudes are invariant under parity (spatial reflection), which is
realized by $w_i\leftrightarrow\bar w_i$ in the multi-Regge
limit~\cite{Prygarin:2011gd}, that is
$x\leftrightarrow\bar x$ and $y\leftrightarrow\bar y$. Secondly, the
multi-Regge limit amplitude should be invariant under
target-projectile symmetry (exchange of the two ingoing momenta), which
amounts to symmetry under $w_5\leftrightarrow
1/w_6$~\cite{Bartels:2014ppa}, that is $x\leftrightarrow y$ and $\bar
x\leftrightarrow\bar y$. The sum
of six-point terms that is subtracted in the
definition~\eqref{eq:g3def} of the function $g_3$ is
separately invariant under these transformations, and hence we can
require parity as well as target-projectile symmetry for the
function~$g_3$ by itself. These symmetries significantly reduce the
number of free parameters in the components of $g_3$, as summarized
in~\tabref{tab:g3params}. In particular, target-projectile symmetry
also swaps $\varepsilon_5$ and $\varepsilon_6$, such that it fixes
$g_3^{0,1}$ and $h_3^{0,1}$ uniquely in terms of~$g_3^{1,0}$ and $h_3^{1,0}$.

Both parity and target projectile symmetry are not trivially
implemented: The parity map replaces all holomorphic weight vectors
and arguments of our single-valued basis functions $\Gs$ with their
complex conjugates. Again using the antipode, these conjugate
single-valued functions can be re-expressed in terms of single-valued
functions with holomorphic arguments~\cite{DelDuca:2016lad}, but those
will not necessarily be elements of the basis~\eqref{eq:Gsbasis}.
Similarly, the target-projectile inversion map $x\leftrightarrow y$
produces non-basis functions. In order to derive constraints for our
ansatz coefficients, all non-basis functions need to be re-expressed
in terms of basis functions, which is possible due to the many
relations among multiple polylogarithms, such as shuffle and stuffle
algebra relations. The single-valued map~\cite{DelDuca:2016lad} is an algebra
homomorphism, hence every identity among ordinary multiple
polylogarithms lifts to a corresponding identity among single-valued
multiple polylogarithms. In this way, single-valued multiple
polylogarithms inherit the shuffle and stuffle algebra relations from
their ordinary counterparts, as well as the simpler rescaling property
\begin{equation}
\Gs(a_1,\dots,a_n;z)=\Gs(ca_1,\dots,ca_n;cz)
\qquad\text{for }
a_n\neq0\text{ and }c\neq0\,.
\label{eq:rescaleG}
\end{equation}
While one can relate non-basis functions back to basis functions by
suitably combining the right shuffle and stuffle identities, it is
often more direct to just match a non-basis function to a
combination of basis functions using numerics, at least up to the
relatively low weight that we consider here. For example, all multiple
polylogarithms up to weight four can be expressed in terms of
classical polylogarithms $\Li_m(z)$ as well as $\Li_{2,2}(z)$ using
e.g.~the \mathematica\ package provided by~\cite{Frellesvig:2016ske}.
Since classical polylogarithms can be readily evalueated numerically,
it is straightforward to match all non-basis functions up to weight
four against combinations of basis functions and zeta values.
However, numerics beyond weight four are
not readily available, and thus implementing target-projectile
symmetry for $g_{3,\mathrm{R}}^{0,0}$ would require to compile all
function identities at weight
five by algebraic means. We have not attempted to do so, as it is
rather laborious, and looking at~\tabref{tab:g3params},
target-projectile symmetry for $g_{3,\mathrm{R}}^{0,0}$ would yield
around $\sim\!15$ more constraints, which would not get us
significantly closer to determining the function $g_3$ completely.
Parity invariance is less demanding in that regard, as the only
functions that cannot be related back to basis functions by simple
shuffle algebra relations are harmonic polylogarithms of weight four
or less. In~\appref{app:polylogrelations}, we list some of the
relations among single-valued polylogarithms that are needed to
evaluate parity and target-projectile symmetry, and we provide all
further relations in an ancillary file.

\paragraph{Collinear Limit.}

Another set of constraints comes from the expansion around the
collinear limit. Since the BDS amplitude correctly captures the
leading behavior in the collinear limit in the Mandelstam regions that
we consider, the remainder function has to vanish in this limit.
In order to take the collinear limit, we map our variables $(x,\bar
x)$ and $(y,\bar y)$ back to $F_{5,6}$ and $r_{5,6}$ via~\eqref{eq:xy}
and~\eqref{eq:frtow7}, which gives
\begin{alignat}{2}
x &= -\frac{F_5 r_5 (1 + F_6 r_6)}{F_6 r_6}
\,,&\qquad
\bar x &= -{r_5 (F_6 + r_6)}{F_5 r_6}
\,,\\
y &= -\frac{F_6 (1 + F_5 r_5) r_6}{F_5 r_5}
\,,&
\bar y &= -\frac{(F_5 + r_5) r_6}{F_6 r_5}
\,.
\end{alignat}
While the Regge limit sits at $T_j\to0$, $S_j\to\infty$ with
$r_j=1/(S_jT_j)$ fixed, the collinear limit is defined by $T_j\to0$
with $S_j$ finite. From the Regge limit, the combined Regge-collinear
limit is therefore attained by letting $r_j\to\infty$, that is
\begin{equation}
x\approx-F_5r_5\to-\infty
\,,\qquad
y\approx-F_6r_6\to-\infty
\,.
\label{eq:xycoll}
\end{equation}
In this limit, the harmonic polylogarithm part
\begin{equation}
\bigbrc{G(\vec a,x)\,|\,a_i\in\brc{0,1}}
\cup\brc{\text{c.c.}}
\,,
\end{equation}
of the basis~\eqref{eq:Gbasis} expands into logarithms and inverse
powers of $x$ (and $\bar x$). For the other part of the basis:
\begin{equation}
\bigbrc{G(\vec a,1/y)\,|\,a_i\in\brc{0,1,x}}
\cup\brc{\text{c.c.}}
\,,
\end{equation}
the expansion is even simpler, since the argument $1/y$ tends to zero,
while the weights $x$ tend to infinity. The basis functions expand to
\begin{align}
G(0,1/y)&=\log(1/y)
\,,\\
G(0,\dots,0,1,1/y)&=-1/y+\order{1/y^2}
\,,\\
G(0,\dots,0,x,1/y)&=-1/xy+\order{1/y^2}
\,,
\end{align}
and all functions $G(\dots,1/y)$ with more than two non-zero weights
are of order $\order{1/y^2}$. After writing all single-valued
functions $\Gs$ in terms of ordinary multiple
polylogarithms and applying the above, we obtain the
expansions of the ansätze
near the collinear limit~\eqref{eq:xycoll}. In doing so, one has to be
careful in picking consistent branches for all occurring logarithms.
Every single-valued multiple polylogarithm
expands to a power series in $\log(r_i)$ and $1/r_i$, where the series
coefficients are rational functions of $F_5$ and $F_6$ as well as zeta values.

The first constraint comes from the fact that the remainder function
should vanish in the collinear limit, that
is there should be no terms that are free of $1/r_i$ factors. This
already implies 37 further constraints on the ansatz, as can be seen
in the fifth line in~\tabref{tab:g3params}. Moreover, we can require
consistency with the general form of the Wilson loop OPE that governs
the remainder function in the collinear
limit~\cite{Basso:2013vsa,Basso:2013aha}. The general systematics of the
Wilson loop OPE predicts that the remainder function in the combined
Regge-collinear limit (at three loops and in any kinematic region)
takes the form
\begin{align}
R_7^{\mathrm{MRL-coll}}
&=\frac{\cos(\phi_5)}{r_5}f_5\bigbrk{\log(\varepsilon_6),\log(r_5)}
+\frac{\cos(\phi_6)}{r_6}f_6\bigbrk{\log(\varepsilon_5),\log(r_6)}
\nn\\
&+\frac{\cos(\phi_5+\phi_6)}{r_5r_6}
h\bigbrk{\log(\varepsilon_5),\log(\varepsilon_6),\log(r_5),\log(r_6)}
\nn\\
&+\frac{\cos(\phi_5-\phi_6)}{r_5r_6}
\bar h\bigbrk{\log(\varepsilon_5),\log(\varepsilon_6),\log(r_5),\log(r_6)}
+\order{r_5^{-2}}+\order{r_6^{-2}}
\,,
\label{eq:opeform}
\end{align}
where $F_i=e^{i\phi_i}$, and $f_5$, $f_6$, $h$, and $\bar h$ are
polynomials in the respective logarithms. In particular, the dependence on $\phi_5$ and
$\phi_6$ is very restricted.%
\footnote{The form~\eqref{eq:opeform} is valid in the Euclidean region
as well as the $(\p\p\p)$ region. During the analytic continuation
into the $(\m\m\m)$ region, all cross ratios $U_{ij}$ follow closed
loops with identical start and end points. Moreover, in the
Basso--Sever--Vieira expressions for the cross ratios in general
kinematics~\cite{Basso:2015rta}, $\phi_5$ and $\phi_6$ only appear in
the combinations $\cos(\phi_5)$, $\cos(\phi_6)$, and
$\cos(\phi_5+\phi_6)$. The cosine is an entire function, and hence the
general form~\eqref{eq:opeform} is preserved under the analytic
continuation into the $(\m\m\m)$ region.}
A general combination of multiple
polylogarithms would also produce sine functions of $\phi_5$,
$\phi_6$, and $\phi_5\pm\phi_6$. It turns out that our parity and
target-projectile symmetric ansatz is already free of such sine terms,
which is an important cross-check of our result. Moreover, terms where
$\cos(\phi_5)$ multiplies $\log(\varepsilon_5)$ or $\log(r_6)$ should
be absent, and the same is true for products of $\cos(\phi_6)$ with
$\log(\varepsilon_6)$ or with $\log(r_5)$.%
\footnote{Note the flipping of the indices $5$ and $6$ in~\eqref{eq:frtow7}.}
The absence of such terms provides yet more constraints on the
coefficients in our ansatz for $g_3$, as can be seen in the sixth line
in~\tabref{tab:g3params}.

When considering the above constraints,
one has to keep in mind that the remainder function in the $(\m\m\m)$
region consists of the function $g_3$ as well as two copies of the
six-point $(\m\m)$ region remainder function~\eqref{eq:3loopallterms}.
The six-point three-loop remainder function in multi-Regge kinematics
has been determined in~\cite{Dixon:2011pw,Dixon:2013eka}.
In principle, there could be cross-terms between the six-point
functions and the function $g_3$, such that only their sum vanishes
and satisfies~\eqref{eq:opeform} in the collinear limit. However, we
have checked that all coefficients (LLA, NLLA, and NNLLA, real and
imaginary parts) of the six-point function
separately vanish and satisfy~\eqref{eq:opeform} in the seven-point
Regge-collinear limit, for both arguments $v_5=-x$ and $v_6=-y$. Hence
also $g_3$ has to satisfy these constraints by itself.

\bigskip

\noindent
In fact, the Regge-collinear limit is not unique: By cyclically
rotating the tessellation of the heptagon that defines the OPE
variables~\eqref{eq:opevars} and taking appropriate limits in the
variables $S_i$, we can probe different limits in the space of
multi-Regge kinematics. Not all collinear limits have an overlap with
the multi-Regge limit: The requirement is that the vanishing of
``small'' cross ratios $u_{j,2}$, $u_{j,3}$ is compatible with the
collinear limit $T_5,T_6\to0$. One further case where this is
satisfied is the cyclic rotation of the Basso--Sever--Vieira variables
by $4$ sites, that is we use the momentum twistors
$Z\suprm{here}_i=Z\suprm{BSV}_{i+4}$, where $Z\suprm{BSV}_i$ are
defined in Appendix~A of~\cite{Basso:2013aha}.%
\footnote{In the case considered above, we used
$Z_i=Z\suprm{BSV}_{i+1}$.}
In this case, the collinear-Regge limit is attained by setting
$S_5=r_5T_5$, $S_6=1/(r_6T_6)$, and letting $T_5,T_6\to0$. The
multi-Regge parameters $w_5$, $w_6$ are then related to the OPE
variables by
\begin{equation}
r_5^2=\frac{1}{w_5\bar w_5}
\,,\qquad
r_6^2=\frac{1}{w_6\bar w_6}
\,,\qquad
F_5^2=\frac{w_5}{\bar w_5}
\,,\qquad
F_6^2=\frac{w_6}{\bar w_6}
\,,
\label{eq:rFwshifted}
\end{equation}
which implies
\begin{equation}
x=-\frac{F_6(F_5+r_5)}{r_5r_6}
\,,\qquad
y=-\frac{r_5(F_6+r_6)}{F_5F_6}
\,.
\end{equation}
Conversely, the combined collinear-Regge limit is attained from the
multi-Regge limit by inverting~\eqref{eq:rFwshifted} for $w_5$, $w_6$,
and letting $r_5,r_6\to\infty$, which implies
\begin{equation}
x\approx-\frac{F_6}{r_6}\to0
\,,\qquad
y\approx-\frac{r_5r_6}{F_5F_6}\to-\infty
\,.
\end{equation}
In this case, the expansion of the basis functions~\eqref{eq:Gbasis}
is even simpler, since all arguments $x$, $\bar x$, $1/y$, and
$1/\bar y$ tend to zero. Expanding the ansätze for our component
functions, we can again require \emph{(i)} vanishing of all components
in the collinear limit, and \emph{(ii)} agreement with the general
form~\eqref{eq:opeform} of the Wilson loop OPE. These constraints
further reduce the ansätze by a few parameters, as shown in the last
two lines of~\tabref{tab:g3params}. It turns out that this second collinear
limit does not yield new further constraints at LLA and NLLA. The
NNLLA functions on the other hand do get constrained further.

\paragraph{The Final Answer.}

Putting all pieces together, one arrives at the most general
combination of multiple polylogarithms that is parity symmetric,
target-projectile symmetric, agrees with the symbol of $g_3$, vanishes
in the collinear limit, and matches the general form of the Wilson
loop OPE in the collinear limit. The resulting function is too bulky
for display here, but is attached in the \mathematica\ file
\filename{g3fctn.m}. It still contains $155$ undetermined
coefficients, as summarized in~\tabref{tab:g3params}. The space of
parameters could perhaps be further reduced by matching subleading
terms in the expansion in $1/r_i$ around the collinear limit to the
predictions from the Wilson loop OPE~\cite{Basso:2013aha}, or by
inspecting the double discontinuity of the symbol. We defer a
more detailed analysis of these further constraints to future work.

The functions $g_3^{1,1}$ and $h_3^{1,1}$ constitute the LLA part of
the function $g_3$, they solely stem from the first diagram
in~\eqref{eq:g3def}, and are not affected by the subtraction of the
six-point functions $f_{1,(3)}$. We display the full LLA part of $g_3$
in~\appref{app:g3lla}. The NLLA parts $g_3^{1,0}$ and $h_3^{1,0}$
comprise the diagrams $2$, $4$, and $7$ in~\eqref{eq:g3def}, and the
functions $g_3^{0,1}$, and $h_3^{0,1}$ consist of the diagrams $3$,
$5$, and $6$.
Finally,
$g_3^{0,0}$ and $h_3^{0,0}$ constitute the NNLLA part of the function
$g_3$, and are composed of
the last eight diagrams in~\eqref{eq:g3def}.
Notably, the NLLA parts $g_3^{1,0}$ and $h_3^{1,0}$ of transcendental
weight four are expressed
solely in terms of products of lower-weight functions $\Gs$. That is,
they are free of weight-four functions $\Gs$ that cannot be expressed
in terms of lower-weight functions. This is a general feature of
single-valued multiple polylogarithms: Every real even-weight
single-valued multiple polylogarithm is expressible in terms of
(products of) lower-weight functions. This property follows from the
defining map~\eqref{eq:svmap} together with the fact that complex
conjugation acts on single-valued functions through the antipode map.%
\footnote{I thank the JHEP referee for pointing out this fact.}
%

\section{Conclusion}
\label{sec:concl}

\paragraph{Summary.}

Exponentiation and factorization are core features of the Regge limit.
In the expansion around large logarithms, they admit a reconstruction
of perturbative amplitudes to any multiplicity, once the BFKL building
blocks (eigenvalues, impact factors, emission blocks) are known. In
this work, we have made this reconstruction explicit, up to the
three-loop order.

A central result is the
relation~\eqref{eq:3loopallterms}, which expresses the simplest cut
contribution to the $n$-point remainder function at three loops in
terms of a few basic building blocks. It should be emphasized that the
identity has a two-fold meaning: On the one hand, it holds at the
level of the complete remainder function's \emph{symbol}. On the
other hand, it holds at the level of full
\emph{functions} once one restricts the remainder function to its
simplest cut contribution as in~\eqref{eq:ncut}, neglecting the Regge
pole terms as well as higher Regge cut contributions such as the ones
in the last line of~\eqref{eq:allcuts}.
The decomposition of the two-Reggeon cut contribution into building
blocks is closely tied to the map~\eqref{eq:vmap} between conventional
multi-Regge limit variables $w_i$ and ``building-block variables''
$v_i$.

The second main result is the determination of the three-loop building
block $g_3$ at the level of the symbol from the known seven-point
three-loop symbol for general kinematics.
Together with the symbol of the known six-point building
block~\cite{Dixon:2011pw}, this permits the reconstruction of the
three-loop remainder function symbol at leading logarithmic order, as
implemented in the attached \mathematica\ file.

Finally, we have constructed a function representative for the
building block $g_3$, based on the knowledge of its symbol as well as the
relevant function space, and by imposing further constraints such as
parity invariance, target-projectile symmetry, and consistency with
the Wilson loop OPE.

\paragraph{Outlook.}

It would be interesting to better understand the general relation
between the BFKL building blocks---impact factors, eigenvalues, and
emission blocks---and the perturbative building blocks that we found
for the full cut contributions. Of course, this relation is in
principle provided by the Fourier--Mellin transform. However, the
action of the inverse Fourier--Mellin transform on general expressions
of multiple polylogs is (to the author's knowledge) not understood
systematically. Especially, it would be interesting to understand how
much can be learnt about the BFKL building blocks when the cut
contributions are only known at the symbol level.
A better understanding of this point would admit to
extract the NLO emission block from two-loop data, from which the
three-loop NLO building blocks $g_L$ and $g_R$ could then be constructed.

We have only fully determined the three-loop seven-point
building block $g_3$ at leading functional transcendental weight. The
parts with lower functional weight (which are multiplied by $\pi$ and zeta
values) have been constrained by symmetry requirements, but still
contain considerable uncertainty in the form of unfixed rational
coefficients. It would be desirable to further constrain the space of
parameters, for example by a more detailed comparison to the Wilson loop
OPE~\cite{Basso:2013aha}. This would require to explicitly compute the
functions $f_5$, $f_6$, $h$, and $\bar h$ by taking the relevant
discontinuity of the three-loop OPE answer. In fact, Basso,
Caron-Huot and Sever could extract the two-particle cut $f_{k,(1)}$ to
all loop orders from the six-point Wilson loop OPE by an ingenuous
analytic continuation in the spectral parameter plane\cite{Basso:2014pla}. Of course, reconstructing the full
three-loop cut contribution $f_{k,(3)}$ for any number of points also
requires knowledge of the
higher building blocks $g\subrm{L}$, $g\subrm{R}$ (at NLLA), and $h$
(at NNLLA). Beyond that,
constructing the full multi-Regge limit remainder function at subleading
functional transcendentality in all kinematic regions also requires to take
more general multi-Reggeon cut terms into account, such as the ones shown
in the last line of~\eqref{eq:allcuts}. While it is possible to
project out these more general cut terms by restricting to kinematic
regions where only adjacent momenta have been flipped,
these higher cut terms form an interesting subject on their own,
and it would be very interesting to understand them systematically.

\subsection*{Acknowledgments}

I sincerely wish to thank Jochen Bartels for many very instructive and enjoyable
discussions, as well as for comments on the manuscript. I also want to thank
Johannes Br\"odel, Vsevolod Chestnov, Georgios Papathanasiou, Volker Schomerus, and Martin
Sprenger for valuable discussions. My work is supported by a Marie
Curie International Outgoing Fellowship within the 7$^{\mathrm{th}}$
European Community Framework Programme under Grant
No.~PIOF-GA-2011-299865.

\appendix

\section{Reduction Identities}
\label{app:reduction-identities}

Here, we want to derive the reduction
identities~\eqref{eq:ifpvreduction}. The absorption of adjacent
emission blocks into impact factors was demonstrated
in~\cite{Bartels:2011ge}, and the analysis directly implies the
reduction identity for emission blocks alone. We reproduce it here for
completeness.

\begin{figure}
\centering
\includegraphics{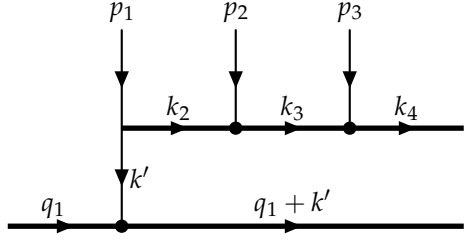}
\caption{Leading-order impact factor and emission vertices. The total
$t$-channel momenta are denoted $q_i$ (see~\figref{fig:kinematics}), hence the
momentum on the lower Reggeon line equals $q_1+k'=q_2-k_2=q_3-k_3=q_4-k_4$.}
\label{fig:reduction}
\end{figure}

The identities are most easily understood in momentum space. At
leading order, the central emission block simply consists of a single
effective Reggeon-Reggeon-gluon vertex attached to the upper Reggeon
line in the two-Reggeon state, see~\figref{fig:reduction}. For a produced
gluon with definite helicity, this effective vertex
equals~\cite{Lipatov:1991nf}
\begin{equation}
\includegraphicsbox{FigRRGvertextree}
=
-\sqrt{2}\,\frac{k_2\bar k_3}{\bar p_2}
\,.
\end{equation}
Compared to the full amplitude, the remainder function has the tree
amplitude divided out. We are computing cut contributions to the
remainder function, hence we need to divide by the tree-level
expression for gluon emission
\begin{equation}
-\sqrt{2}\,\frac{q_2\bar q_3}{\bar p_2}
\,.
\end{equation}
At leading order, the central emission
block in momentum space therefore equals
\begin{equation}
C_2\equiv C(q_2,k_2,p_2)
=
\frac{k_2\bar k_3}{q_2\bar q_3}
=
\frac{\bar k_2\brk{k_2+p_2}}{\bar q_2\brk{q_2+p_2}}
\,.
\label{eq:Cmom}
\end{equation}
Combining two leading-order emission blocks requires to include the
intermediate transverse propagator $1/\abs{k_3}^2$, again divided by
the corresponding tree-level expression $1/\abs{q_3}^2$:
\begin{equation}
C_2\cdot
\frac{\abs{q_3}^2}{\abs{k_3}^2}\cdot
C_3
=
\frac{k_2\bar k_3}{q_1\bar q_3}
\cdot
\frac{\abs{q_3}^2}{\abs{k_3}^2}
\cdot
\frac{k_3\bar k_4}{q_3\bar q_4}
=
\frac{k_2\bar k_4}{q_2\bar q_4}
=
\frac{\bar k_2\brk{k_2+p_2+p_3}}{\bar q_2\brk{q_2+p_2+p_3}}
\,.
\end{equation}
This clearly equals the single emission block~\eqref{eq:Cmom} with the emitted
momentum $p_2$ replaced by the sum of momenta $p_2+p_3$. Iterating the
procedure straightforwardly yields the reduction
identity for emission blocks, on the right in~\eqref{eq:ifpvreduction}.

The leading-order impact factor consists of a single gluon emission
from the bottom Reggeon line, as in~\figref{fig:reduction}. It thus
reads~\cite{Bartels:2008sc}
\begin{equation}
-\sqrt{2}\,\frac{q_1(\bar q_2-\bar k_2)}{\bar k'}
=
-\sqrt{2}\,\frac{q_1\bar q_2}{\bar p_1}
-\sqrt{2}\,\frac{\abs{q_1}^2\bar k_2}{\bar k'\bar p_1}
\,.
\end{equation}
On the right, the emission factor has been split into a ``local''
piece (first term) and a ``non-local'' part (second term). The
local piece plays a role for the one-loop amplitude, but does not
affect the remainder function~\cite{Bartels:2008sc,Bartels:2011ge}. It
therefore can be dropped for the purpose of computing discontinuities of
the remainder function. Thus only the second term
remains for the leading-order impact factor. Dividing by the tree
expression $-\sqrt{2}\,q_1\bar q_2/\bar p_1$, it becomes
\begin{equation}
\Phi_{\mathrm{L},1}\equiv\Phi\subrm{L}(q_1,k',p_1)
=
\frac{\bar q_1\bar k_2}{\bar k'\bar q_2}
=
\frac{\bar q_1\brk{\bar k'+\bar p_1}}{\bar k'\brk{\bar q_1+\bar p_1}}
\,.
\label{eq:ifmom}
\end{equation}
Combining this impact factor with an adjacent emission block, one
again needs to include the intermediate propagator factor, which yields
\begin{equation}
\Phi_{\mathrm{L},1}\cdot
\frac{\abs{q_2}^2}{\abs{k_2}^2}\cdot
C_2
=
\frac{\bar q_1\bar k_2}{\bar k'\bar q_2}
\cdot
\frac{\abs{q_2}^2}{\abs{k_2}^2}
\cdot
\frac{k_2\bar k_3}{q_2\bar q_3}
=
\frac{\bar q_1\bar k_3}{\bar k'\bar q_3}
=
\frac{\bar q_1\brk{\bar k'+\bar p_1+\bar p_2}}{\bar k'\brk{\bar q_1+\bar p_1+\bar p_2}}
\,.
\end{equation}
This equals the original impact factor~\eqref{eq:ifmom} with the emitted momentum $p_1$ replaced
by the sum of momenta $p_1+p_2$. Again, iterating the procedure yields
the reduction identity for impact factors, on the left
in~\eqref{eq:ifpvreduction}.
%

\section{Four-Loop Expansion}

At four loops, the two-Reggeon cut at LLA evidently expands to a sum of
six-point, seven-point, and eight-point functions:
\begin{multline}
f_{k,(4)}
=\sum_{j=5}^{k+4}f_{1,(4)}(\varepsilon_j;v_{5,k+4;j})
+\sum_{i=5}^{k+3}\sum_{j=i+1}^{k+4}g_{2,(4)}(\varepsilon_i,\varepsilon_j;v_{5,j-1;i},v_{i+1,k+4;j})
\\
+\sum_{i=5}^{k+2}\sum_{j=i+1}^{k+3}\sum_{m=j+1}^{k+4}g_{3,(4)}(\varepsilon_i,\varepsilon_j,\varepsilon_m;v_{5,j-1;i},v_{i+1,m-1;j},v_{j+1,k+4;m})
+\order{\mathrm{NLLA}}
\,,
\label{eq:fk4LLA}
\end{multline}
where
\begin{equation}
g_{2,(4)}(\varepsilon_5,\varepsilon_6;w_5,w_6)
\equiv
 f_{2,(4)}(\varepsilon_5,\varepsilon_6;w_5,w_6)
-f_{1,(4)}(\varepsilon_5;v_{5,6;5})
-f_{1,(4)}(\varepsilon_6;v_{5,6;6})
\,,
\end{equation}
and
\begin{align}
g_{3,(4)}(\varepsilon_5,\varepsilon_6,\varepsilon_7;w_5,w_6,w_7)
\equiv\mspace{-180mu}&
\nn\\&
 f_{3,(4)}(\varepsilon_5,\varepsilon_6,\varepsilon_7;w_5,w_6,w_7)
-f_{1,(4)}(\varepsilon_5;v_{5,7;5})
-f_{1,(4)}(\varepsilon_6;v_{5,7;6})
-f_{1,(4)}(\varepsilon_7;v_{5,7;7})
\nn\\&
-g_{2,(4)}(\varepsilon_5,\varepsilon_6;w_5,v_{6,7;6})
-g_{2,(4)}(\varepsilon_5,\varepsilon_7;v_{5,6;5},v_{6,7;7})
-g_{2,(4)}(\varepsilon_6,\varepsilon_7;v_{5,6;6},w_7)
\,.
\end{align}
At NLLA, there are three more building blocks that all stem from
nine-point data:
\newcommand{\dg}[2]{%
\text{\small$\setlength{\arraycolsep}{0.5pt}\biggsbrk{\begin{matrix}#1\\[-2pt]#2\end{matrix}}$}}
\begin{align}
g_{4,(4),1}(\varepsilon_5,\varepsilon_6,\varepsilon_7,\varepsilon_8;w_5,w_6,w_7,w_8)
&=\dg{1&&1&&1&&1&&1}{0&1&0&1&0&0&1&0&0}
 -\dg{1&&1&&1&&2}{0&1&0&1&0&0&1}
 -\dg{1&&1&&2&&1}{0&2&0&0&1&0&0}
 \label{eq:g41}
 \\&\mspace{-220mu}
 -\dg{1&&1&&2&&1}{0&1&0&1&1&0&0}
 -\dg{1&&1&&2&&1}{0&1&0&0&1&1&0}
 +\dg{1&&1&&3}{0&2&0&0&1}
 +\dg{1&&3&&1}{0&2&1&0&0}
 +\dg{1&&1&&3}{0&1&0&1&1}
 -\dg{1&&4}{0&2&1}
 +\order{\text{NNLLA}}\,,
 \nn
\\
g_{4,(4),2}(\varepsilon_5,\varepsilon_6,\varepsilon_7,\varepsilon_8;w_5,w_6,w_7,w_8)
&=\dg{1&&1&&1&&1&&1}{0&1&0&0&1&0&0&1&0}
 +\dg{1&&3&&1}{0&1&1&1&0}
 +\order{\text{NNLLA}}\,,
\\
g_{4,(4),3}(\varepsilon_5,\varepsilon_6,\varepsilon_7,\varepsilon_8;w_5,w_6,w_7,w_8)
&=\dg{1&&1&&1&&1&&1}{0&0&1&0&0&1&0&1&0}
 -\dg{2&&1&&1&&1}{1&0&0&1&0&1&0}
 -\dg{1&&2&&1&&1}{0&0&1&0&0&2&0}
 \label{eq:g43}
 \\&\mspace{-220mu}
 -\dg{1&&2&&1&&1}{0&0&1&1&0&1&0}
 -\dg{1&&2&&1&&1}{0&1&1&0&0&1&0}
 +\dg{3&&1&&1}{1&0&0&2&0}
 +\dg{1&&3&&1}{0&0&1&2&0}
 +\dg{3&&1&&1}{1&1&0&1&0}
-\dg{4&&1}{1&2&0}
 +\order{\text{NNLLA}}\,.
 \nn
\end{align}
Here, each bracket stands for a BFKL diagram, with the following
notation: In the bottom sequence, the numbers alternatingly stand for
impact factors\,/\,emission blocks and BFKL Green's functions. For the
impact factors and emission blocks, the number specifies the loop
order. For the Green's functions, it specifies the number of (leading
order) BFKL eigenvalues, accompanied by the respective large logarithm
$\log(\varepsilon_j)$; i.e. the numbers $1$, $2$, and $3$ stand for
the first, second, and third terms in~\eqref{eq:gfexpansion}. The
numbers in the top row specify how many momenta are attached to the
respective impact factor or emission block. The
expansion~\eqref{eq:fk4LLA} of the two-Reggeon cut extends to
\begin{multline}
f_{k,(4)}
=f_{1,(4)}^{\brc{i_1,i_2}}
+g_{2,(4)}^{\brc{i_1,i_2,i_3}}
+g_{3,(4)}^{\brc{i_1,i_2,i_3,i_4}}
\\
+g_{4,(4),1}^{\brc{i_1,i_2,i_3,1,i_4}}
+g_{4,(4),2}^{\brc{i_1,i_2,1,i_3,i_4}}
+g_{4,(4),3}^{\brc{i_1,1,i_2,i_3,i_4}}
+\order{\mathrm{NNLLA}}
\,.
\end{multline}
Here, each term stands for a sum over partitions
$\brc{i_1,i_2,\dots}$, $\sum_j i_j=k+1$,
of the external momenta $p_4,\dots,p_{k+4}$ into subsequences
$\brk{p_4,\dots,p_{3+i_1}}$, $\brk{p_{4+i_1},\dots,p_{3+i_1+i_2}}$,
\dots, whose sums get attached to the momentum slots of the respective
building block. For example,
\begin{equation}
g_{4,(4),1}^{\brc{i_1,i_2,i_3,1,i_4}}
\equiv
\sum_{i=5}^{k+1}\sum_{j=i+1}^{k+2}\sum_{m=j+2}^{k+4}
g_{4,(4),1}(\varepsilon_i,\varepsilon_j,\varepsilon_{m-1},\varepsilon_m;v_{5,j-1;i},v_{i+1,m-2;j},v_{j+1,m-1;m-1},v_{m,k+4;m})
\,.
\end{equation}
At NNLLA and NNNLLA, there are many more terms and building blocks.
The complete expansion of the two-Reggeon cut reads
\begin{align}
f_{k,(4)}={}&
 f_{1,(4)}^{\brc{i_1,i_2}}
+g_{2,(4)}^{\brc{i_1,i_2,i_3}}
+g_{3,(4)}^{\brc{i_1,i_2,i_3,i_4}}
+g_{4,(4),1}^{\brc{i_1,i_2,i_3,1,i_4}}
+g_{4,(4),2}^{\brc{i_1,i_2,1,i_3,i_4}}
+g_{4,(4),3}^{\brc{i_1,1,i_2,i_3,i_4}}
\nn\\&
+g_{4,(4),4}^{\brc{1,i_1,1,i_2,i_3}}
+g_{4,(4),5}^{\brc{1,i_1,i_2,1,i_3}}
+g_{4,(4),6}^{\brc{i_1,1,1,i_2,i_3}}
+g_{4,(4),7}^{\brc{i_1,1,i_2,1,i_3}}
+g_{4,(4),8}^{\brc{i_1,1,i_2,i_3,1}}
\nn\\&
+g_{4,(4),9}^{\brc{i_1,i_2,1,1,i_3}}
+g_{4,(4),10}^{\brc{i_1,i_2,1,i_3,1}}
+g_{5,(4),1}^{\brc{i_1,1,i_2,1,i_3,i_4}}
+g_{5,(4),2}^{\brc{i_1,1,i_2,i_3,1,i_4}}
+g_{5,(4),3}^{\brc{i_1,i_2,1,i_3,1,i_4}}
\nn\\&
+g_{4,(4),11}^{\brc{i_1,1,i_2,1,1}}
+g_{4,(4),12}^{\brc{i_1,1,1,i_2,1}}
+g_{4,(4),13}^{\brc{i_1,1,1,1,i_2}}
+g_{4,(4),14}^{\brc{1,i_1,1,i_2,1}}
+g_{4,(4),15}^{\brc{1,i_1,1,1,i_2}}
+g_{4,(4),16}^{\brc{1,1,i_1,1,i_2}}
\nn\\&
+g_{5,(4),4}^{\brc{i_1,1,i_2,1,i_3,1}}
+g_{5,(4),5}^{\brc{i_1,1,i_2,1,1,i_3}}
+g_{5,(4),6}^{\brc{i_1,1,1,i_2,1,i_3}}
+g_{5,(4),7}^{\brc{1,i_1,1,i_2,1,i_3}}
+g_{6,(4)}^{\brc{i_1,1,i_2,1,i_3,1,i_4}}
\,.
\end{align}
Terms in the second and third lines start at $\order{\text{NNLLA}}$,
terms in the last two lines are of order $\order{\text{NNNLLA}}$.
The individual terms are listed explicitly in the following. The
completion of the NLLA terms~\eqref{eq:g41}-\eqref{eq:g43} is given by
\setcounter{MaxMatrixCols}{14}
\renewcommand{\dg}[2]{%
\text{\scriptsize$\setlength{\arraycolsep}{0.5pt}\Bigsbrk{\begin{matrix}#1\\[-2pt]#2\end{matrix}}$}}
\setlength{\multlinegap}{0pt}
\begin{flalign}
g_{4,(4),1}\mspace{-40mu}&\mspace{40mu}
=\dg{1&&1&&1&&1&&1}{0&1&0&1&0&0&1&0&0}
-\dg{1&&1&&1&&2}{0&1&0&1&0&0&1}
-\dg{1&&1&&2&&1}{0&2&0&0&1&0&0}
-\dg{1&&1&&2&&1}{0&1&0&1&1&0&0}
-\dg{1&&1&&2&&1}{0&1&0&0&1&1&0}
+\dg{1&&1&&3}{0&2&0&0&1}
+\dg{1&&3&&1}{0&2&1&0&0}
&\nn\\&
+\dg{1&&1&&3}{0&1&0&1&1}
-\dg{1&&4}{0&2&1}
-\dg{2&&1&&1&&1}{1&0&0&0&1&1&0}
-\dg{2&&1&&1&&1}{1&0&0&1&1&0&0}
-\dg{1&&1&&1&&2}{0&0&1&0&0&1&1}
-\dg{1&&1&&1&&2}{0&1&0&0&1&0&1}
-\dg{1&&1&&1&&2}{1&0&0&1&0&0&1}
-\dg{1&&1&&2&&1}{0&0&1&0&1&1&0}
&\nn\\&
-\dg{1&&1&&2&&1}{0&0&1&1&1&0&0}
-\dg{1&&1&&2&&1}{0&1&0&0&1&0&1}
-\dg{1&&1&&2&&1}{0&1&0&0&2&0&0}
-\dg{1&&1&&2&&1}{0&1&1&0&1&0&0}
-\dg{1&&1&&2&&1}{1&0&0&0&1&1&0}
-\dg{1&&1&&2&&1}{1&0&0&1&1&0&0}
-\dg{1&&1&&2&&1}{1&1&0&0&1&0&0}
&\nn\\&
-\dg{1&&1&&2&&1}{0&(1)&0&0&1&0&0}
+\dg{1&&1&&3}{1&0&0&1&1}
+\dg{1&&1&&3}{1&1&0&0&1}
+\dg{1&&1&&3}{0&0&1&1&1}
+\dg{1&&1&&3}{0&1&1&0&1}
+\dg{1&&1&&3}{0&1&0&0&2}
+\dg{1&&1&&3}{0&(1)&0&0&1}
-\dg{2&&3}{1&1&1}
-\dg{1&&1&&1&&2}{0&0&1&0&0&0&2}
\mspace{-100mu}
&\nn\\&
-\dg{1&&1&&2&&1}{1&0&0&0&2&0&0}
-\dg{1&&1&&2&&1}{1&0&0&0&1&0&1}
+\dg{1&&1&&3}{0&0&1&0&2}
+\dg{1&&1&&3}{0&0&2&0&1}
+\dg{1&&1&&3}{1&0&0&0&2}
+\dg{1&&1&&3}{2&0&0&0&1}
+\dg{1&&1&&3}{1&0&1&0&1}
-\dg{2&&3}{1&0&2}
&
\end{flalign}
\begin{flalign}
g_{4,(4),2}\mspace{-40mu}&\mspace{40mu}
=\dg{1&&1&&1&&1&&1}{0&1&0&0&1&0&0&1&0}
+\dg{1&&3&&1}{0&1&1&1&0}
+\dg{1&&3&&1}{1&0&1&1&0}
+\dg{1&&3&&1}{1&1&1&0&0}
+\dg{1&&3&&1}{0&0&1&1&1}
+\dg{1&&3&&1}{0&1&1&0&1}
+\dg{1&&3&&1}{0&0&2&1&0}
+\dg{1&&3&&1}{0&1&2&0&0}
&\nn\\&
+\dg{1&&3&&1}{0&0&1&(1)&0}
+\dg{1&&3&&1}{0&(1)&1&0&0}
+\dg{1&&4}{0&1&2}
+\dg{1&&4}{0&(1)&1}
+\dg{1&&4}{1&1&1}
+\dg{4&&1}{1&1&1}
+\dg{4&&1}{1&(1)&0}
+\dg{4&&1}{2&1&0}
+\dg{1&&3&&1}{0&0&1&0&2}
+\dg{1&&3&&1}{2&0&1&0&0}
&\nn\\&
+\dg{1&&3&&1}{0&0&2&0&1}
+\dg{1&&3&&1}{1&0&2&0&0}
+\dg{1&&3&&1}{0&0&3&0&0}
+\dg{1&&3&&1}{1&0&1&0&1}
-\dg{1&&4}{0&0&3}
-\dg{1&&4}{1&0&2}
-\dg{1&&4}{2&0&1}
-\dg{4&&1}{1&0&2}
-\dg{4&&1}{2&0&1}
-\dg{4&&1}{3&0&0}
\mspace{-20mu}
&
\end{flalign}
\begin{flalign}
g_{4,(4),3}\mspace{-40mu}&\mspace{40mu}
=\dg{1&&1&&1&&1&&1}{0&0&1&0&0&1&0&1&0}
-\dg{2&&1&&1&&1}{1&0&0&1&0&1&0}
-\dg{1&&2&&1&&1}{0&0&1&0&0&2&0}
-\dg{1&&2&&1&&1}{0&0&1&1&0&1&0}
-\dg{1&&2&&1&&1}{0&1&1&0&0&1&0}
+\dg{3&&1&&1}{1&0&0&2&0}
+\dg{1&&3&&1}{0&0&1&2&0}
&\nn\\&
+\dg{3&&1&&1}{1&1&0&1&0}
-\dg{4&&1}{1&2&0}
-\dg{1&&1&&1&&2}{0&1&1&0&0&0&1}
-\dg{1&&1&&1&&2}{0&0&1&1&0&0&1}
-\dg{2&&1&&1&&1}{1&1&0&0&1&0&0}
-\dg{2&&1&&1&&1}{1&0&1&0&0&1&0}
-\dg{2&&1&&1&&1}{1&0&0&1&0&0&1}
-\dg{1&&2&&1&&1}{0&1&1&0&1&0&0}
&\nn\\&
-\dg{1&&2&&1&&1}{0&0&1&1&1&0&0}
-\dg{1&&2&&1&&1}{1&0&1&0&0&1&0}
-\dg{1&&2&&1&&1}{0&0&2&0&0&1&0}
-\dg{1&&2&&1&&1}{0&0&1&0&1&1&0}
-\dg{1&&2&&1&&1}{0&1&1&0&0&0&1}
-\dg{1&&2&&1&&1}{0&0&1&1&0&0&1}
-\dg{1&&2&&1&&1}{0&0&1&0&0&1&1}
&\nn\\&
-\dg{1&&2&&1&&1}{0&0&1&0&0&(1)&0}
+\dg{3&&1&&1}{1&1&0&0&1}
+\dg{3&&1&&1}{1&0&0&1&1}
+\dg{3&&1&&1}{1&1&1&0&0}
+\dg{3&&1&&1}{1&0&1&1&0}
+\dg{3&&1&&1}{2&0&0&1&0}
+\dg{3&&1&&1}{1&0&0&(1)&0}
-\dg{3&&2}{1&1&1}
-\dg{2&&1&&1&&1}{2&0&0&0&1&0&0}
\mspace{-40mu}
&\nn\\&
-\dg{1&&2&&1&&1}{0&0&2&0&0&0&1}
-\dg{1&&2&&1&&1}{1&0&1&0&0&0&1}
+\dg{3&&1&&1}{2&0&1&0&0}
+\dg{3&&1&&1}{1&0&2&0&0}
+\dg{3&&1&&1}{2&0&0&0&1}
+\dg{3&&1&&1}{1&0&0&0&2}
+\dg{3&&1&&1}{1&0&1&0&1}
-\dg{3&&2}{2&0&1}
&
\end{flalign}
Here, a $(1)$ in the place of a Green's function stands for the
one-loop correction to the BFKL eigenvalue, i.e.\ for the fourth term
in~\eqref{eq:gfexpansion}.
The following terms start at $\order{\text{NNLLA}}$:
\begin{flalign}
g_{4,(4),4}={}
&\dg{1&&1&&1&&1&&1}{1&0&0&0&1&0&0&1&0}
&&
\\
g_{4,(4),5}={}
&\dg{1&&1&&1&&1&&1}{1&0&0&1&0&0&1&0&0}
-\dg{1&&1&&2&&1}{2&0&0&0&1&0&0}
-\dg{1&&1&&2&&1}{1&0&1&0&1&0&0}
-\dg{1&&1&&2&&1}{0&0&2&0&1&0&0}
-\dg{2&&1&&1&&1}{1&0&0&0&2&0&0}
-\dg{3&&2}{1&0&2}
&&
\\
g_{4,(4),6}={}
&\dg{1&&1&&1&&1&&1}{0&0&1&0&1&0&0&1&0}
-\dg{1&&1&&2&&1}{0&0&1&0&1&0&1}
-\dg{1&&1&&2&&1}{0&0&1&0&2&0&0}
-\dg{2&&1&&1&&1}{1&0&1&0&1&0&0}
-\dg{2&&1&&1&&1}{1&0&1&0&0&0&1}
&&
\\
g_{4,(4),7}={}
&\dg{1&&1&&1&&1&&1}{0&0&1&0&0&0&1&1&0}
+\dg{1&&1&&1&&1&&1}{0&0&1&0&0&1&1&0&0}
+\dg{1&&1&&1&&1&&1}{0&0&1&1&0&0&1&0&0}
+\dg{1&&1&&1&&1&&1}{0&1&1&0&0&0&1&0&0}
+\dg{2&&2&&1}{1&0&1&1&0}
+\dg{2&&2&&1}{1&1&1&0&0}
+\dg{2&&1&&2}{1&0&0&1&1}
\mspace{-20mu}
&&\nn\\&
+\dg{2&&1&&2}{1&1&0&0&1}
+\dg{1&&2&&2}{0&0&1&1&1}
+\dg{1&&2&&2}{0&1&1&0&1}
+\dg{1&&1&&1&&1&&1}{0&0&1&0&0&0&2&0&0}
+\dg{1&&1&&1&&1&&1}{0&0&2&0&0&0&1&0&0}
+\dg{2&&2&&1}{2&0&1&0&0}
+\dg{2&&2&&1}{1&0&2&0&0}
&&\nn\\&
+\dg{2&&2&&1}{1&0&1&0&1}
+\dg{2&&1&&2}{1&0&0&0&2}
+\dg{2&&1&&2}{2&0&0&0&1}
+\dg{1&&2&&2}{0&0&1&0&2}
+\dg{1&&2&&2}{0&0&2&0&1}
+\dg{1&&2&&2}{1&0&1&0&1}
&&
\\
g_{4,(4),8}={}
&\dg{1&&1&&1&&1&&1}{0&0&1&0&0&1&0&0&1}
-\dg{1&&2&&1&&1}{0&0&1&0&0&0&2}
-\dg{1&&2&&1&&1}{0&0&1&0&1&0&1}
-\dg{1&&2&&1&&1}{0&0&1&0&2&0&0}
-\dg{1&&1&&1&&2}{0&0&2&0&0&0&1}
-\dg{2&&3}{2&0&1}
&&
\\
g_{4,(4),9}={}
&\dg{1&&1&&1&&1&&1}{0&1&0&0&1&0&1&0&0}
-\dg{1&&2&&1&&1}{1&0&1&0&1&0&0}
-\dg{1&&2&&1&&1}{0&0&2&0&1&0&0}
-\dg{1&&1&&1&&2}{0&0&1&0&1&0&1}
-\dg{1&&1&&1&&2}{1&0&0&0&1&0&1}
&&
\\
g_{4,(4),10}={}
&\dg{1&&1&&1&&1&&1}{0&1&0&0&1&0&0&0&1}
&&
\end{flalign}
\begin{flalign}
g_{5,(4),1}&
=\dg{1& &1& &1& &1& &1& &1}{0&0&1&0&0&0&1&0&0&1&0}
+\dg{2&&2&&1&&1}{1&0&1&0&0&1&0}
&
\\
g_{5,(4),2}&
=\dg{1&&1&&1&&1&&1&&1}{0&0&1&0&0&1&0&0&1&0&0}
+\dg{2&&1&&1&&2}{1&0&0&1&0&0&1}
+\dg{1&&2&&1&&2}{0&0&1&0&0&1&1}
+\dg{3&&1&&1&&1}{1&0&0&0&1&1&0}
+\dg{3&&1&&1&&1}{1&0&0&1&1&0&0}
+\dg{1&&1&&1&&3}{0&0&1&1&0&0&1}
+\dg{1&&1&&1&&3}{0&1&1&0&0&0&1}
\mspace{-200mu}
&
\nn\\
&
+\dg{2&&1&&2&&1}{1&1&0&0&1&0&0}
+\dg{1&&2&&2&&1}{0&0&1&0&1&1&0}
+\dg{1&&2&&2&&1}{0&0&1&1&1&0&0}
+\dg{1&&2&&2&&1}{0&1&1&0&1&0&0}
+\dg{3&&3}{1&1&1}
-\dg{1&&2&&1&&2}{0&0&1&0&2&0&0}
-\dg{2&&1&&2&&1}{0&0&2&0&1&0&0}
\mspace{-200mu}
&
\\
g_{5,(4),3}&
=\dg{1&&1&&1&&1&&1&&1}{0&1&0&0&1&0&0&0&1&0&0}
+\dg{1&&1&&2&&2}{0&1&0&0&1&0&1}
&
\end{flalign}
Finally, the following terms only contribute at NNNLLA:
\begin{flalign}
g_{4,(4),11}&=
\dg{1&&1&&1&&1&&1}{0&0&1&0&0&0&1&0&1}
-\dg{2&&1&&1&&1}{1&0&0&0&1&0&1}
&
g_{4,(4),12}&=
\dg{1&&1&&1&&1&&1}{0&0&1&0&1&0&0&0&1}
&
\\
g_{4,(4),13}&=
 \dg{1&&1&&1&&1&&1}{0&0&1&0&1&0&1&0&0}
+\dg{2&&1&&2}{1&0&1&0&1}
&
g_{4,(4),14}&=
\dg{1&&1&&1&&1&&1}{1&0&0&0&1&0&0&0&1}
&
\\
g_{4,(4),15}&=
\dg{1&&1&&1&&1&&1}{1&0&0&0&1&0&1&0&0}
&
g_{4,(4),16}&=
 \dg{1&&1&&1&&1&&1}{1&0&1&0&0&0&1&0&0}
-\dg{1&&1&&1&&2}{1&0&1&0&0&0&1}
&
\end{flalign}
\begin{flalign}
g_{5,(4),4}&=
\dg{1&&1&&1&&1&&1&&1}{0&0&1&0&0&0&1&0&0&0&1}
&
g_{5,(4),5}&=
\dg{1&&1&&1&&1&&1&&1}{0&0&1&0&0&0&1&0&1&0&0}
\mspace{150mu}
&
\\
g_{5,(4),6}&=
\dg{1&&1&&1&&1&&1&&1}{0&0&1&0&1&0&0&0&1&0&0}
&
g_{5,(4),7}&=
\dg{1&&1&&1&&1&&1&&1}{1&0&0&0&1&0&0&0&1&0&0}
&
\end{flalign}
\begin{flalign}
g_{6,(4)}&
=\dg{1&&1&&1&&1&&1&&1&&1}{0&0&1&0&0&0&1&0&0&0&1&0&0}
-\dg{1&&2&&2&&2}{0&0&1&0&1&0&1}
-\dg{2&&2&&2&&1}{1&0&1&0&1&0&0}
&
\end{flalign}
The large number of terms at NNLLA and NNNLLA shows that the
decomposition into building blocks is less effective than at two and
three loops. The reason is that the reduction
identities~\eqref{eq:ifpvreduction} are only established for the
leading order emission block. It would be interesting to see whether
the (thus far unknown) NLO emission block satisfies similar reduction
identities, in which case many of the above terms could be
reduced and absorbed into a smaller number of building blocks, reducing the
complexity of the decomposition.

\section{Polylogarithm Identities}
\label{app:polylogrelations}

When applying the target-projectile transformation $x\leftrightarrow
y$ or the parity map $x\to\bar x$, $y\to\bar y$ to the ansatz
functions for the components of $g_3$, some single-valued basis
functions are mapped to non-basis single-valued multiple polylogarithms. In order to
impose the required symmetries on the ansätze, these non-basis functions need to
be re-expressed in terms of basis functions. This can always be
achieved with the help of shuffle and stuffle relations, as well as
the rescaling property~\eqref{eq:rescaleG}. Shuffle relations take the
form
\begin{equation}
G(\vec a;z)G(\vec b;z)
=\sum_{\vec c\,\in\,\vec a\shuffle\vec b}G(\vec c;z)\,,
\label{eq:shuffle}
\end{equation}
where the sum runs over all shuffles $\vec a\shuffle\vec b$ of the
weight vectors $\vec a$ and $\vec b$, that is over all permutations of
their components that preserve the ordering of elements within both
$\vec a$ and $\vec b$. The shuffle relation follows directly from the
iterated integral definition of multiple
polylogarithms~\eqref{eq:Gdef}, and they hold for ordinary as well as
single-valued multiple polylogarithms.
Stuffle relations on the other hand are less transparent in the
integral representation; they follow directly from the series
representation of $G(\vec a;z)$ around $z=0$, see for
example~\cite{Duhr:2014woa}.

At weights one and two, the following identities are needed to
evaluate the parity and target-projectile invariance conditions:
\begin{align}
\Gsshort{y}_{0}
&=-\Gsshort{\cy}_{0}
\,,
&
\Gsshort{\cx}_{0}
&=-\Gsshort{x}_{0}
\,,
&
\Gsshort{1}_{x}
&=-\Gsshort{x}_{0}+\Gsshort{x}_{1}
\,,
\\
\Gsshort{x}_{x\cy}
&=-\Gsshort{\cy}_{0}+\Gsshort{\cy}_{1}
\,,
&
\Gsshort{\cy}_{x\cy}
&=-\Gsshort{x}_{0}+\Gsshort{x}_{1}
\,,
&
\Gsshort{1}_{0,x}
&=\half\brk{\Gsshort{x}_{0}}^2-\Gsshort{x}_{0,1}
\,,
\\
\Gsshort{x}_{0,x\cy}
&=\half\brk{\Gsshort{\cy}_{0}}^2-\Gsshort{\cy}_{0,1}
\,,
&
\Gsshort{\cy}_{0,x\cy}
&=\half\brk{\Gsshort{x}_{0}}^2-\Gsshort{x}_{0,1}
\,,
&&
\\
\Gsshort{\cy}_{x\cy,x}
&=-\Gsshort{x}_{0}\Gsshort{\cy}_{1}+\Gsshort{x}_{1}\Gsshort{\cy}_{1}+\Gsshort{\cy}_{0,x}-\Gsshort{\cy}_{1,x}
\,,
\mspace{-150mu}&&&&
\end{align}
Here and in the following, we use the condensed notation
$\Gsshort{z}_{a_1,\dots,a_n}\equiv\Gs(a_1,\dots,a_n;z)$, and
$\cy\equiv1/y$. At weight three, we have for example the following
identities:
\begin{align}
\Gsshort{x}_{0,0,x\cy}
&=-\sfrac{1}{6}\brk{\Gsshort{\cy}_{0}}^3+\Gsshort{\cy}_{0,0,1}
\,,
\\
\Gsshort{x}_{0,x\cy,x\cy}
&=-\sfrac{1}{6}\brk{\Gsshort{\cy}_{0}}^3+\Gsshort{\cy}_{0}\Gsshort{\cy}_{0,1}-\Gsshort{\cy}_{0,0,1}-\Gsshort{\cy}_{0,1,1}+2\zeta_3
\,,
\\
\Gsshort{\cy}_{0,0,x\cy}
&=-\sfrac{1}{6}\brk{\Gsshort{x}_{0}}^3+\Gsshort{x}_{0,0,1}
\,,
\\
\Gsshort{\cy}_{0,x,x\cy}
&=-\half\brk{\Gsshort{x}_{0}}^2\Gsshort{\cy}_{1}+\Gsshort{\cy}_{1}\Gsshort{x}_{0,1}+\Gsshort{x}_{0}\Gsshort{\cy}_{0,1}-\Gsshort{x}_{1}\Gsshort{\cy}_{0,1}-\Gsshort{x}_{0}\Gsshort{\cy}_{0,x}+\Gsshort{x}_{1}\Gsshort{\cy}_{0,x}
\nn\\&\mspace{20mu}  +\Gsshort{\cy}_{1}\Gsshort{\cy}_{0,x}-2\Gsshort{\cy}_{0,0,x}-\Gsshort{\cy}_{0,x,1}
\,,
\\
\Gsshort{\cy}_{0,x\cy,x}
&=\half\brk{\Gsshort{x}_{0}}^2\Gsshort{\cy}_{1}-\Gsshort{\cy}_{1}\Gsshort{x}_{0,1}-\Gsshort{\cy}_{1}\Gsshort{\cy}_{0,x}+\Gsshort{\cy}_{0,0,x}+\Gsshort{\cy}_{0,1,x}+\Gsshort{\cy}_{0,x,1}
\,,
\\
\Gsshort{\cy}_{0,x\cy,x\cy}
&=-\sfrac{1}{6}\brk{\Gsshort{x}_{0}}^3+\Gsshort{x}_{0}\Gsshort{x}_{0,1}-\Gsshort{x}_{0,0,1}-\Gsshort{x}_{0,1,1}+2\zeta_3
\,,
\\
\Gsshort{\cy}_{x\cy,x,x}
&=-\half\Gsshort{x}_{0}\brk{\Gsshort{\cy}_{1}}^2+\half\Gsshort{x}_{1}\brk{\Gsshort{\cy}_{1}}^2+\Gsshort{\cy}_{1}\Gsshort{\cy}_{0,x}-\Gsshort{\cy}_{0,1,x}-\Gsshort{\cy}_{0,x,1}
\nn\\&\mspace{20mu}
+\Gsshort{\cy}_{0,x,x}-\Gsshort{\cy}_{1,1,x}-\Gsshort{\cy}_{1,x,x}
\,,
\\
\Gsshort{\cy}_{x\cy,x\cy,x}
&=\half\brk{\Gsshort{x}_{0}}^2\Gsshort{\cy}_{1}-\Gsshort{x}_{0}\Gsshort{x}_{1}\Gsshort{\cy}_{1}+\half\brk{\Gsshort{x}_{1}}^2\Gsshort{\cy}_{1}+\half\Gsshort{x}_{0}\brk{\Gsshort{\cy}_{1}}^2-\half\Gsshort{x}_{1}\brk{\Gsshort{\cy}_{1}}^2
\nn\\&\mspace{20mu}
-\Gsshort{x}_{0}\Gsshort{\cy}_{0,1}+\Gsshort{x}_{1}\Gsshort{\cy}_{0,1}-\Gsshort{\cy}_{1}\Gsshort{\cy}_{0,x}+\Gsshort{\cy}_{0,0,x}+\Gsshort{\cy}_{0,x,1}+\Gsshort{\cy}_{1,1,x}
\,.
\end{align}
All of the above equations rely on multiple shuffle and stuffle
relations. These are all relations involving both letters $x$ and $y$
that one needs for parity and target-projectile symmetry up to
weight four. These as well as all further required relations among
single-valued harmonic polylogarithms only involving $x$ can be found
in the ancillary file \filename{GGtobasis.m}.

\section{The Function \texorpdfstring{$g_3$}{g3} at LLA}
\label{app:g3lla}

For reference, we display the LLA part of the function $g_3$, up to
a few undetermined coefficients that are not constrained by the
symmetries that we considered:%
\footnote{This result agrees with~\cite{DelDuca:2016lad} (equation
(D.8) there) once we set all parameters $c\suprm{i/r}_{..}$ to zero.}
\begin{align}
g_3^{1,1}& + 2\pi i\,h_3^{1,1}=
\frac{1}{16}\Bigbrk{
\brk{\Gsshort{x}_{0}}^2\Gsshort{\cy}_{1}
-2\Gsshort{x}_{0}\Gsshort{\cy}_{0}\Gsshort{\cy}_{1}
+2\Gsshort{\cy}_{0}\Gsshort{x}_{1}\Gsshort{\cy}_{1}
-\brk{\Gsshort{x}_{1}}^2\Gsshort{\cy}_{1}
\nn\\&
+\Gsshort{x}_{0}\brk{\Gsshort{\cy}_{1}}^2
-\Gsshort{x}_{1}\brk{\Gsshort{\cy}_{1}}^2
+2\Gsshort{\cy}_{0}\Gsshort{x}_{1}\Gsshort{\cy}_{x}
+\brk{\Gsshort{x}_{1}}^2\Gsshort{\cy}_{x}
-2\Gsshort{\cy}_{0}\Gsshort{\cy}_{1}\Gsshort{\cy}_{x}
\nn\\&
-2\Gsshort{x}_{1}\Gsshort{\cy}_{1}\Gsshort{\cy}_{x}
+\brk{\Gsshort{\cy}_{1}}^2\Gsshort{\cy}_{x}
-2\Gsshort{\cy}_{x}\Gsshort{x}_{0,1}
+2\Gsshort{\cy}_{x}\Gsshort{1,\cy}_{0}
+2\Gsshort{\cy}_{1}\Gsshort{\cy}_{0,x}
+2\Gsshort{x}_{0}\Gsshort{x,\cy}_{1}
\nn\\&
-2\Gsshort{x}_{1}\Gsshort{\cy}_{1,x}
+2\Gsshort{\cy}_{1}\Gsshort{x,\cy}_{1}
-4\Gsshort{\cy}_{0,1,x}
-4\Gsshort{\cy}_{0,x,1}
-4\Gsshort{\cy}_{1,1,x}}
+c\suprm{r}_{2,1}\Gsshort{\cy}_{x}\zeta_2
\nn\\&
+2\pi i\Bigbrk{
c\suprm{i}_{2,1}\bigbrk{\brk{\Gsshort{x}_{0}}^2
-2\Gsshort{x}_{0}
\Gsshort{x}_{1}
+\brk{\Gsshort{x}_{1}}^2
+\brk{\Gsshort{\cy}_{1}}^2}
+c\suprm{i}_{2,2}\bigbrk{\Gsshort{x}_{0}-\Gsshort{x}_{1}}\Gsshort{\cy}_{1}
\nn\\&
+c\suprm{i}_{2,3}\bigbrk{\Gsshort{x}_{0}
-\Gsshort{x}_{1}
-\Gsshort{\cy}_{1}}\Gsshort{\cy}_{x}
+c\suprm{i}_{2,4}\bigbrk{\Gsshort{\cy}_{0}
-\Gsshort{x}_{1}
-\Gsshort{\cy}_{1}}\Gsshort{\cy}_{x}
+c\suprm{i}_{2,5}\brk{\Gsshort{\cy}_{x}}^2}
\,.
\end{align}
Here, we again used the shorthand notation
$\Gsshort{z}_{a_1,\dots,a_n}\equiv\Gs(a_1,\dots,a_n;z)$. The NLLA and
NNLLA functions are too lengthy for display, they are attached in the
ancillary file \filename{g3fcn.m}.

\bibliographystyle{nb}
\bibliography{references}

\end{document}